\documentclass{emulateapj}

\usepackage{graphicx}
\usepackage[space]{grffile}
\usepackage{latexsym}
\usepackage{amsfonts,amsmath,amssymb}
\usepackage{url}
\usepackage[utf8]{inputenc}
\usepackage{fancyref}
\usepackage{hyperref}
\hypersetup{colorlinks=false,pdfborder={0 0 0},}
\usepackage{textcomp}
\usepackage{longtable}
\usepackage{multirow,booktabs}

\usepackage{natbib}

\usepackage{supertabular}

\usepackage[caption=false]{subfig}

\def\mps{m\,s$^{-1}$~}
\def\mpsp{m\,s$^{-1}$}
\def\cmps{cm\,s$^{-1}$~}

\def\ha{H$\alpha$~}
\def\hap{H$\alpha$}
\def\shk{$S_{HK}$~}
\def\shkp{$S_{HK}$}
\def\id{$I_{\textrm{D}}$~}

\def\iha{$I_{\textrm{H}\alpha}$~}
\def\ihap{$I_{\textrm{H}\alpha}$}
\def\kone{\ion{K}{1}~}

\def\naone{\ion{Na}{1}~}

\def\catwo{\ion{Ca}{2}~}

\def\wk{$W_{\textrm{K\,I}}$~}
\def\wkp{$W_{\textrm{K\,I}}$}

\def\wnap{$W_{\textrm{Na\,I NIR}}$}
\def\wnad{$W_{\textrm{Na\,I D}}$~}

\def\wcaone{$W_{\textrm{Ca\,I 8498}}$~}
\def\wcaonep{$W_{\textrm{Ca\,I 8498}}$}
\def\wcatwo{$W_{\textrm{Ca\,I 8542}}$~}
\def\wcatwop{$W_{\textrm{Ca\,I 8542}}$}
\def\wcathree{$W_{\textrm{Ca\,I 8662}}$~}

\slugcomment{Accepted for publication in The Astrophysical Journal (ApJ)}

\shorttitle{}
\shortauthors{Robertson et al.}

\begin{document}


\title{Proxima Centauri as a Benchmark for Stellar Activity Indicators in the Near Infrared}


\author{Paul Robertson$^{1,2,3}$}

\author{Chad Bender$^{2,3}$}

\author{Suvrath Mahadevan$^{2,3}$}

\author{Arpita Roy$^{2,3}$}

\author{Lawrence W. Ramsey$^{2,3}$}

\altaffiltext{1}{NASA Sagan Fellow}
\altaffiltext{2}{Department of Astronomy and Astrophysics, The Pennsylvania State University}
\altaffiltext{3}{Center for Exoplanets \& Habitable Worlds, The Pennsylvania State University}

\begin{abstract}

A new generation of dedicated Doppler spectrographs will attempt to detect low-mass exoplanets around mid-late M stars at near infrared (NIR) wavelengths, where those stars are brightest and have the most Doppler information content.  A central requirement for the success of these instruments is to properly measure the component of radial velocity (RV) variability contributed by stellar magnetic activity and to account for it in exoplanet models of RV data.  The wavelength coverage for many of these new instruments will not include the \catwo H\&K or \ha lines, the most frequently used absorption-line tracers of magnetic activity.  Thus, it is necessary to define and characterize NIR activity indicators for mid-late M stars in order to provide simultaneous activity metrics for NIR RV data.  We have used the high-cadence UVES observations of the M5.5 dwarf Proxima Centauri from Fuhrmeister et al.~(2011) to compare the activity sensitivity of 8 NIR atomic lines to that of \hap.  We find that equivalent width-type measurements of the NIR \kone doublet and the \catwo NIR triplet are excellent proxies for the canonical optical tracers.  The \catwo triplet will be acquired by most of the new and upcoming NIR Doppler spectrographs, offering a common, reliable indicator of activity.

\end{abstract}



\bibliographystyle{apj}

\section{\bf Introduction}

The monitoring of time-dependent stellar features produced by magnetic activity in nearby stars is a critical and valuable component of radial velocity (RV) surveys for exoplanets.  RV programs probe stars at temporal resolutions matching a number of stellar phenomena, from rotation (days) to spot lifetimes (months) and magnetic cycles (years).  Some of the most valuable contributions to the observational study of stellar magnetic activity have come from long-term RV surveys \citep[e.g.][]{wright05,if10,lovis11,robertson13}.

In addition to facilitating better understanding of stellar astrophysics, the study of magnetic activity in RV surveys is also vital for identifying and mitigating astrophysical RV noise caused by activity.  Examples abound for systems in which activity-induced RV contributions have obscured \citep{howard13,pepe13,haywood14} or mimicked \citep{queloz01,santos14,robertson14,johnson16} the signals of exoplanets.  Furthermore, as shown by every known observational metric, the problem of astrophysical RV noise, if unaddressed, can only be expected to get worse as high-precision Doppler spectrometers move past the 1 \mps threshold.  For example, \citet{bastien14} find that the quietest stars in their flicker-jitter sample have an RV jitter floor of $\sim3$ \mpsp.  \citet{dumusque15} find a solar RV RMS in excess of 50 \cmps from Sun-as-a-star observations using a solar telescope fiber-fed to the HARPS-N spectrograph, a result consistent with RVs of solar spectra reflected off Vesta \citep{haywood16}.  Activity-induced RV jitter is ubiquitous below the 1 \mps level, and is rapidly becoming the chief impediment to the discovery of low-mass exoplanets.  Treatment of activity noise must be a core component of any high-precision Doppler program, a reality that has been acknowledged by the community in recent years \citep{fischer16}.

A number of new and upcoming precision Doppler spectrographs, in an effort to discover and characterize exoplanets around mid-late M dwarfs, will operate at red-optical to near-infrared (NIR) wavelengths, where late-type stars are brightest.  Examples of such instruments include CARMENES \citep{quirrenbach14}, HPF \citep{mahadevan14}, SPIRou \citep{artigau14}, and MINERVA-Red \citep{blake15}.  Many types of astrophysical RV noise should be suppressed at NIR wavelengths \citep{marchwinski15}; for example, the amplitudes of starspot signals should be reduced, as the contrast between cool spots and the surrounding photosphere is diminished.  Nevertheless, it will be essential for surveys using these instruments to monitor the activity of their targets, and to remain wary of activity-induced false positive planet detections.  

The most reliable and efficient method for tracking stellar activity in RV surveys is to measure the fluxes in one or more stellar lines that are known to be sensitive to chromospheric magnetic activity.  The ``gold standard" for studying time-variable activity in Sunlike stars, and in particular for optical RV instruments and surveys, has been the \catwo H\&K doublet \citep[e.g.][and references therein]{wilson68,vaughan78,baliunas95,if10}.  Some programs using spectrographs lacking blue wavelength coverage and/or targeting cool stars (M dwarfs and giants) have also frequently used the \ha line \citep{kurster03,hatzes15} and occasionally the \naone D doublet \citep{gds11,robertson15} with success.  When possible, it is ideal to measure more than one such tracer, as any star may exhibit multiple activity signals, each of which may (or may not) appear in a given chromospheric line with varying amplitude \citep[e.g.][]{mortier16}.

Some preliminary efforts to obtain precise RV measurements in the NIR have already been conducted \citep[see][for a brief summary]{bean10}.  However, the programs most closely approaching Doppler precisions sufficient to detect exoplanets \citep[$\sigma_{\textrm{RV}}\sim10$ \mpsp, e.g.][]{reiners09,barnes12,barnes14} have all relied on \ha as their primary line diagnostic for stellar activity.  The wavelength ranges of upcoming spectrographs such as HPF and SPIRou will not include \hap.  Furthermore, as has been repeatedly demonstrated for optical RV spectrographs, it is beneficial to obtain multiple spectral activity indicators across the entire bandpass of the instrument.  Therefore, it is imperative for the success of the coming generation of NIR RV instruments to identify and characterize spectral tracers sensitive to the subtle magnetic variations that can create false-positive Doppler signals.

The task of characterizing one or more spectral line activity tracers in the NIR is potentially a major observational undertaking.  One would need to observe a star exhibit some time-dependent variability, and examine how that variability manifests in the flux of each candidate line.  Old, relatively quiet M dwarfs--such as those that will constitute many of the best targets for NIR Doppler surveys--often vary slowly, with rotation periods in excess of 100 days \citep{robertson14,robertson15}.  Thus, the time baseline for an observational campaign would need to be long.  Ideally, the observations would also cover a broad wavelength range, so that the new tracers can be compared to well established lines such as \catwo H\&K and \hap.

For M stars, there is one publicly available data set that potentially obviates the need for an expensive observational effort: the multi-wavelength archive of spectroscopy of Proxima Centauri from UVES and HARPS.  Proxima Centauri (= GJ 551, hereafter Proxima) is an M5.5 dwarf, and the closest known hydrogen-burning star to the Sun.  Proxima was recently shown by \citet{ae16} to host a low-mass exoplanet in its liquid-water habitable zone, further increasing the star's value as an observational target.  The ESO archive\footnote{Based on data obtained from the ESO Science Archive Facility under request numbers 202455, 202466, 203085, and 227038.  Based on observations made with ESO Telescopes at the La Silla Paranal Observatory under programme IDs 071.C-0498, 072.C-0488, 072.C-0495, 078.C-0829, 082.C-0718, 082.D-0953, 173.C-0606, 183.C-0437, and 191.C-0505.} contains 261 optical spectra of Proxima from the HARPS spectrograph \citep{pepe00}, and 722 optical and NIR spectra from the UVES spectrograph \citep{dekker00}.  These spectra will be described fully in Section \ref{sec:data}.  Although neither data set has optimal time sampling, Proxima is highly variable on short time scales, providing a large ``activity baseline" across which to compare spectral lines.

We sought to evaluate potential NIR activity tracers using a bootstrapping strategy.  First, we used the optical spectra to verify that Ca H\&K and \ha are essentially equivalent in terms of sensitivity to activity, with \ha being somewhat preferable due to increased S/N.  Since the NIR UVES observations include \hap, it was then possible to compare the behavior of NIR lines to that of \ha in order to determine the most reliable activity tracers.

We note that while we anticipate these results will be at least partially applicable to all low-mass stars, chromospheric emission of the \kone and \naone lines is only completely collisionally dominated in M dwarf atmospheres \citep{andretta97,takeda02}.  For hotter stars, where the \kone and \naone transitions may become radiation dominated, it is not anticipated that those lines will be as sensitive to chromospheric magnetic activity, although \citet{marchwinski15} did observe and remark on some activity-dependent spectral variability near the optical \naone D doublet regions for the Sun.

The paper is organized thusly.  In Section \ref{sec:data}, we describe the spectra, candidate absorption lines, and activity indices.  Section \ref{sec:telluric} details our correction for telluric absorption near the \kone and \naone NIR doublets.  We briefly discuss the various activity phenomena observed for Proxima in Section \ref{sec:activity}.  In Section \ref{sec:tracers} we establish \catwo H\&K, the \naone D doublet, and \ha as equivalent activity tracers, and compare the candidate NIR tracers' activity sensitivity to that of \hap.  We discuss the results of this experiment in Section \ref{sec:discussion}, and summarize our conclusions in Section \ref{sec:conclusions}.

\section{\bf Data and Definition of Indices} \label{sec:data}

\begin{table*}
\footnotesize
\begin{center}

\begin{tabular}{| l  c | c | c c | c |}
\hline
Species & Wavelength & $W_0$ & $R1$ & $R2$ & Reference \\
 & (\AA) & (\AA) & (\AA) & (\AA) & \\ \hline
\kone & 7664.90 & 1.00 & 7618.5-7620.5 & 7734.0-7736.0 & \multirow{2}{*}{This Work} \\
\kone & 7698.96 & 1.00 & 7618.5-7620.5 & 7734.0-7736.0 &  \\
\naone & 8183.26 & 0.50 & 8140.0-8142.0 & 8206.0-8208.0 & \multirow{2}{*}{Adapted from \citet{schlieder12}} \\
\naone & 8194.82 & 0.50 & 8140.0-8142.0 & 8206.0-8208.0 &  \\
\catwo & 8498.02 & 0.75 & 8474.0-8484.0 & 8560.0-8580.0 & \multirow{3}{*}{Adapted from \citet{cenarro01}} \\
\catwo & 8542.09 & 0.25 & 8474.0-8484.0 & 8560.0-8580.0 &  \\
\catwo & 8662.14 & 0.25 & 8619.0-8642.0 & 8700.0-8725.0 & \\
P$\delta$ & 10049.4 & 0.50 & 10042.5-10047.5 & 10074.0-10079.9 & This Work \\
\hline
\end{tabular}

\caption{\label{tab:boxwidths}
\footnotesize Index measurement windows and reference bands for each of the candidate activity indicator lines considered.}

\end{center}
\end{table*}

\subsection{Data}

A considerable number of time-resolved, high-resolution optical spectra of Proxima are publicly available as a result of Doppler searches for exoplanets in the system.  70 HARPS spectra are available from HARPS exoplanet survey \citep[e.g.][]{pepe04}, in particular the search for exoplanets around nearby M stars \citep{bonfils13}.  Additionally, the ESO archive includes another 191 HARPS spectra taken as part of the Cool Tiny Beats survey, which aims to use high-cadence RV observations to characterize short-term variability of M dwarfs \citep{ae14}.  The HARPS spectrograph provides a resolving power $R\sim115,000$ and wavelength coverage from approximately $3800-6900$\AA.

A second source of optical spectra comes from a survey of exoplanets around M stars conducted with the UVES spectrograph \citep{endl06,endl08,zechmeister09}.  94 spectra were taken using the red arm of UVES with an $0.3\arcsec$ slit, resulting in wavelength coverage from $4950-7040$\AA~and $R = 100,000-120,000$.  A critical difference between the optical HARPS and UVES spectra is that UVES utilizes an iodine (I$_2$) absorption cell as a precision wavelength standard, which superimposes thousands of weak iodine absorption lines over the stellar spectrum from approximately $5000-6000$\AA.  Thus, we do not consider the \naone D doublet from these spectra herein, as they are contaminated by the I$_2$ absorption.

The largest component of our data set is taken from \citet{fuhrmeister11}, who conducted an intensive multi-wavelength monitoring campaign of Proxima.  They observed the star continuously over three nights (9, 11, 13 March 2009) with UVES and \emph{XMM-Newton}, spanning wavelengths from X-rays through the NIR in order to create a physical model of Proxima's flaring chromosphere.

The \citet{fuhrmeister11} observations used a non-standard setup of the red arm to enable wavelength coverage from $6400-10080$\AA, specifically intended to include coverage of \hap.  The spectra have a typical resolving power $R = 45,000$.  The red observations were taken at high cadence, totaling 562 spectra over the three nights.  The data set also includes 66 spectra taken with the blue arm of UVES, providing wavelengths from $3290-4500$\AA, but we focused primarily on the red spectra for this work.

\subsection{Definition of Indices}

Stellar activity is traced via atomic lines by measuring the variable emission from those species in the chromosphere.  For more active M stars like Proxima, certain lines such as Ca H\&K and \ha are dominated by the chromospheric component, leading to the appearance of those lines in emission.  For lines dominated by photospheric absorption, one instead measures the amount of ``filling in" of those lines by chromospheric emission.  In either case, an equivalent width (EW) or similar measurement is appropriate for approximating the state of the line in a given spectrum.

Because the optical and NIR continuum of an M star is completely blanketed by atomic and molecular absorption lines, and because the lines of interest for activity measurement are often located in molecular and/or telluric bands, a true EW is not ideal for the line indices we measure.  Instead, our indices approximate EWs by measuring the flux in a given line weighted by the flux in nearby reference bands, which are selected to avoid large stellar or telluric lines.  Where possible, we have adopted or modified index definitions previously defined in the literature.  Below, we describe our measurements of each of the standard and candidate tracers used.

\begin{figure*}
\begin{center}
\includegraphics[width=1.8\columnwidth]{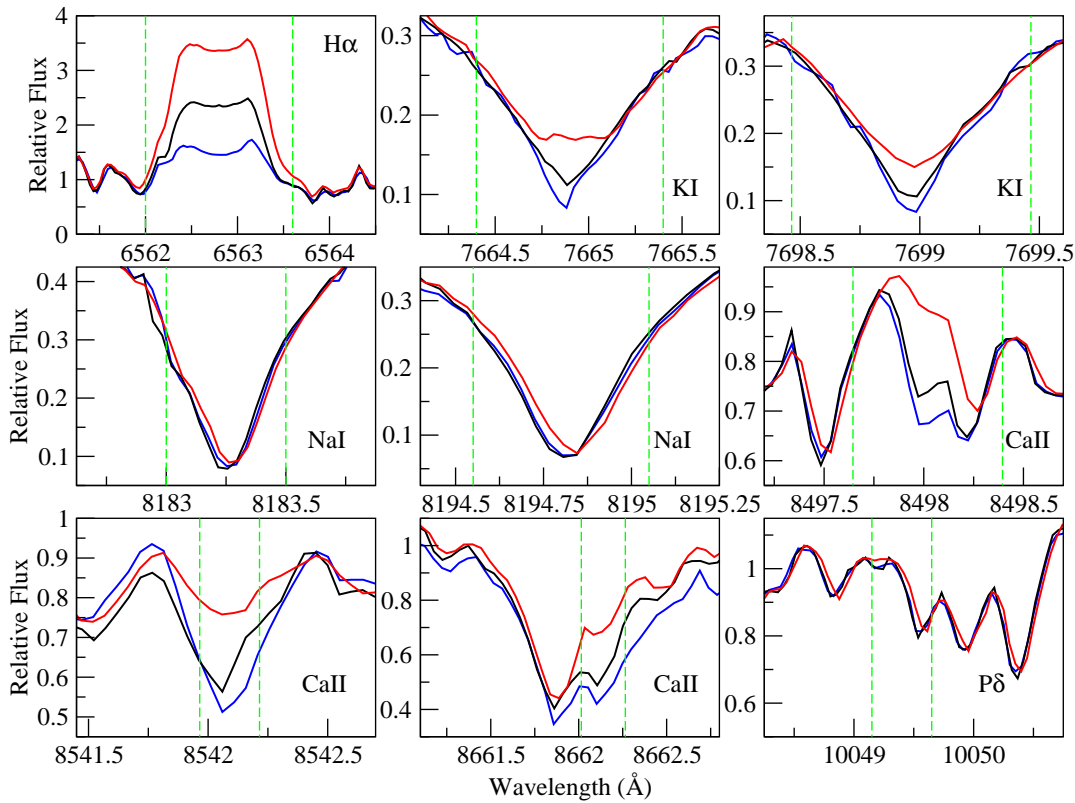}
\caption{\label{fig:lines}
\footnotesize Examples of the lines we have considered as potential NIR activity tracers from the UVES spectra.  Lines are shown at observations of high (\emph{red}) and low (\emph{blue}) relative activity levels, as well as near the average $W$ value for each line (\emph{black}).  \ha is shown for comparison.  The window within which we sum the flux to compute $W$ for each line is bounded by dashed green lines.}
\end{center}
\end{figure*}

\subsubsection{Standard Activity Tracers}

\begin{center}\emph{The \catwo H\&K Lines}\end{center}

The strength of the \catwo H\&K line cores are the field standard activity metric for optical RV surveys.  As is common, we have adopted the Mount Wilson ``S-index" \citep[\shkp;][]{vaughan78} to measure the lines, which is simply the sum of the fluxes inside windows centered on the H ($\lambda = 3968.47$\AA \footnote{Unless indicated otherwise, wavelengths quoted herein are in air.}) and K ($\lambda = 3933.66$\AA) lines weighted by the fluxes within windows on either side of the doublet.

We have adopted the M dwarf-specific definition of \shk from \citet{gds11}.  This definition, itself adapted from that of \citet{boisse09}, uses a $0.6$\AA~window for each line, which is more narrow than is commonly used for hotter stars.

For mid-late M stars such as Proxima, the continuum emission near the H\&K lines is quite weak, making the \shk index less than ideal as an activity tracer.  However, because it is a well-characterized standard in the field of Doppler exoplanet searches, we have included it in our analysis.

\begin{center}\emph{The \naone D Lines}\end{center}

The Na I D doublet ($\lambda\lambda = 5889.96, 5895.93$\AA) is a sensitive indicator of magnetic activity in cool stars, where chromospheric emission of the lines becomes collisionally dominated \citep{diaz07,gds11}.

Because there is no commonly used ``standard" definition of a sodium D index, we have used the same $W$ index described in Section \ref{sec:candidates} below.  Our index \wnad is the average $W$ value for each of the two lines in the doublet.  We have adopted the $0.5$\AA~window size and reference band definitions from \citet{gds11}.

One complication that often arises when using the Na I D lines to measure stellar activity is that the lines may be contaminated by emission lines from telluric sodium \citep{hanuschik03}.  In the case of Proxima, the star's high absolute radial velocity \citep[$v_r = -22.4$ km/s,][]{torres06} shifts the stellar sodium lines enough that the sky lines rarely affect the \wnad measurement.  In general, though, the sky emission lines must be accounted for in order to obtain useful measurements of this activity index.

\begin{center}\emph{The \ha Line}\end{center}

The \ha equivalent width is commonly used to characterize magnetic activity in M stars across a wide range of spectral subtype and mean activity level \citep[e.g.][and references therein]{robertson13,west15}.  It is also the only frequently-used spectral activity tracer common to the blue and red spectra considered herein, making it crucial for evaluating candidate NIR tracers by comparison to established indices.

For Proxima, we measure the \ha index, \ihap, as defined in \citet{robertson13}.  This index is identical to that of \citet{kurster03}, except that--as for \shkp--it uses a more narrow 1.6\AA~window for the flux in the \ha line.  \citet{gds11} first suggested using the narrower window, claiming that the resulting measurements were more tightly correlated with \shkp.

\citet{fuhrmeister11} note that the morphology of the \ha line changes as a function of time.  In particular, the red wing of the line is clearly stronger than the blue wing in many of the UVES spectra. Fuhrmeister et al.~interpret this phenomenon as being caused by material evaporating into the chromosphere during flare events, then raining back down, thus causing blue- and redshifts in the chromospheric lines.  While we agree that these changes in morphology exist, we have not modified \iha to account for line shape changes, and do not observe any peculiar behavior in the index that might be caused by the shape variability.

\subsubsection{Candidate NIR Tracers} \label{sec:candidates}

For each of the candidate line tracers described below, we measure an index analogous to an equivalent width.  In most cases, the lines have at least one property--whether blending, central reversal, or being completely lost within the pseudocontinuum--that makes modeling and integrating a line profile impossible.  Rather, we take the average flux within a band centered on the line, weighted by the average of nearby reference bands chosen to be relatively free of tellurics and molecular bands.  Specifically, our generic line index $W_{X}$ is defined as

\begin{equation}
W_X = W_0\Big{(}1 - \frac{2 \overline{F_X}}{\overline{F_{R1}} + \overline{F_{R2}}}\Big{)}
\end{equation}

\noindent where $W_0$ is the width (in Angstroms) of the band within which the line flux is averaged, $\overline{F_X}$ is the mean flux within that band, and $\overline{F_{R1}}$ and $\overline{F_{R2}}$ are the mean fluxes within each of the two reference bands.  In Table \ref{tab:boxwidths}, we list the values of $W_0$ and the reference bands chosen for each of our candidate lines.  For lines that show activity sensitivity, $W_0$ is chosen so as to maximize the correlation of the index with \ihap.

The measurement uncertainties on $W_{X}$ follow from standard propagation of error, and are defined as

\begin{equation}
\sigma_{W_X} = \frac{W_0 - W_X}{SNR} \sqrt{1 + \frac{\overline{F_{R1}} + \overline{F_{R2}}}{2 \overline{F_X}}}
\end{equation}

\noindent where $SNR$ is the approximate signal-to-noise ratio near the spectral feature being measured.

Below, we briefly introduce the candidate tracers we considered.  In Figure \ref{fig:lines}, we show examples of these lines from UVES spectra at low, average, and high activity levels, with \ha for comparison.

\begin{figure*}
\begin{center}
\subfloat[\label{subfig:terraspec}]{\includegraphics[width=1.2\columnwidth]{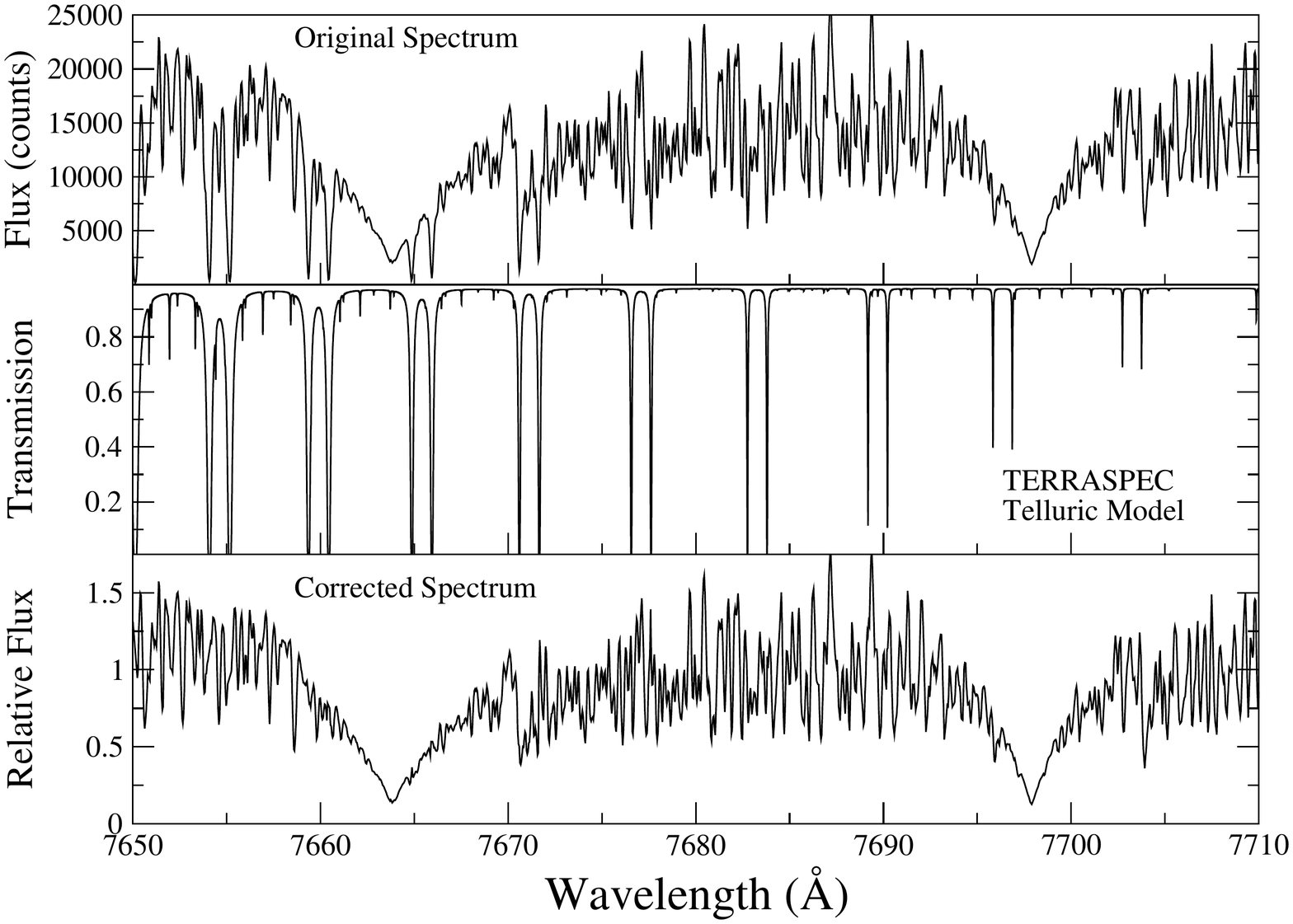}}
\subfloat[\label{subfig:ip}]{\includegraphics[width=0.8\columnwidth]{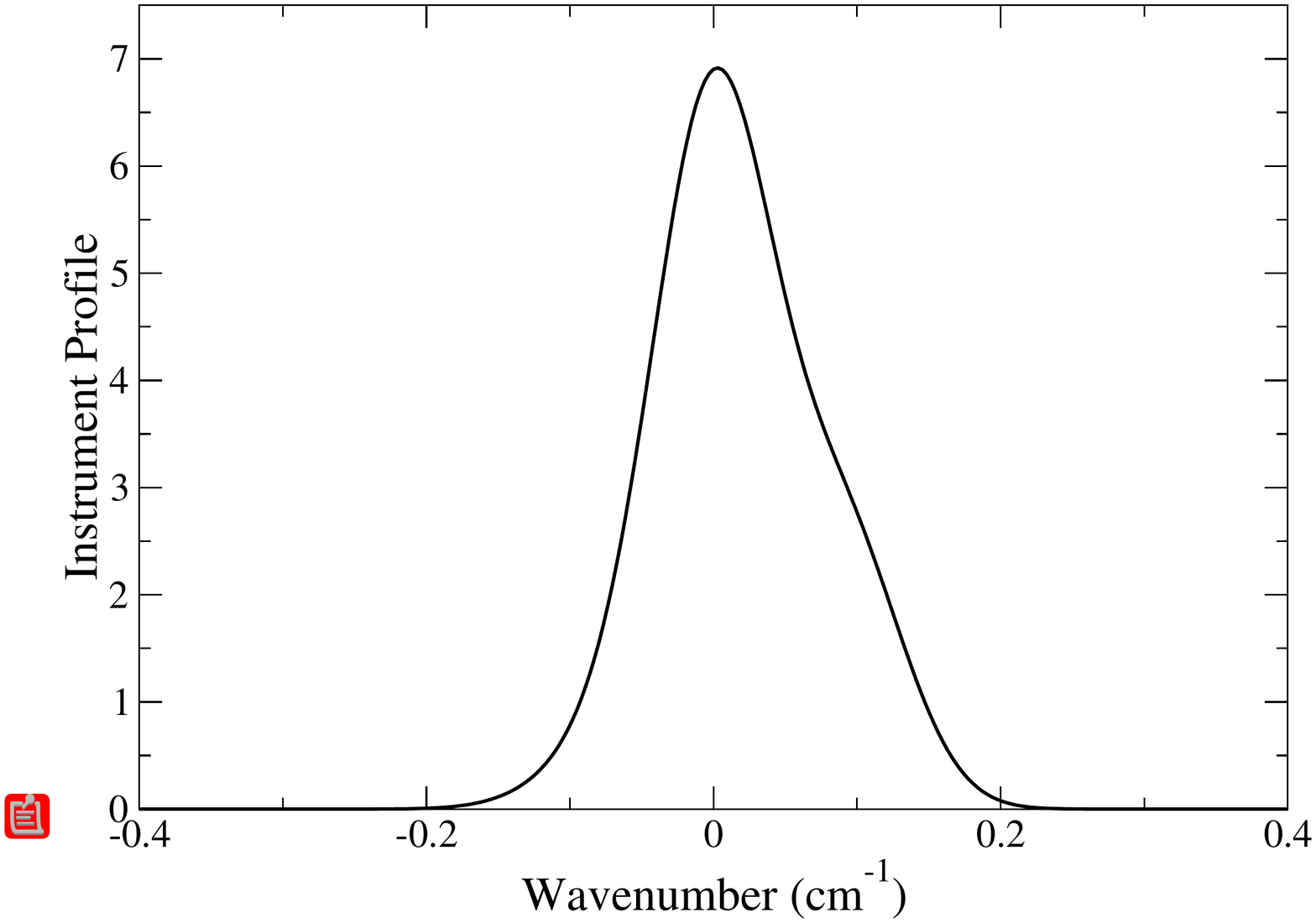}}
\caption{\label{fig:terraspec}
\footnotesize \emph{a}: An example spectrum of Proxima near the \kone doublet before (\emph{top}) and after (\emph{bottom}) correcting for telluric absorption with TERRASPEC.  The computed telluric model is shown in the middle panel.  \emph{b}: The instrument profile (IP), as modeled by TERRASPEC.}
\end{center}
\end{figure*}

\begin{center}\emph{The \catwo NIR Triplet}\end{center}

The \catwo NIR triplet ($\lambda\lambda = 8498.02, 8542.09, 8662.14$\AA) has long been known to be sensitive to chromospheric activity in cool stars \citep{linsky70,mallik97,chmielewski00,andretta05,busa07}.  One or more lines of the triplet are commonly discussed as the tracer most likely to be useful for NIR RV work \citep[e.g.][]{barnes14}.

For Proxima, the line at $8498$\AA~is completely hidden in the pseudocontinuum, and the $8542$\AA~line is weak, occasionally disappearing into the pseudocontinuum as well.  For these reasons, \citet{barnes14} considered only the reddest \catwo line at $8662$\AA, finding a strong correlation with \hap.  For the sake of completeness, we elected to consider all three calcium lines, but because of concerns with line visibility and blending (discussed more later), we evaluated each line separately rather than create a combined index.

\citet{cenarro01} define a generic index for the \catwo NIR triplet, intended to be applicable across a broad range of spectral types.  The windows used for the lines themselves are far too broad to be appropriate for this application, as they are designed to accommodate rapidly rotating stars.  However, we do adopt their reference bands in an effort to make our index definition appropriate across all M spectral types.

\begin{center}\emph{The \naone NIR Doublet}\end{center}

The NIR doublet produced by \naone ($\lambda\lambda = 8183.26, 8194.82$\AA) has a number of features that make it particularly attractive as a potential activity indicator for cool stars.  The doublet is a very strong feature in M star spectra, even at the latest subtypes \citep[see, e.g.][Fig. 3]{wade88}.  Furthermore, while the lines are contaminated by telluric water lines, they sit outside any major molecular bands.

Previous research indicates the NIR sodium doublet is responsive to multiple stellar properties, including activity.  \citet{schlieder12} find that the \naone lines are sensitive to stellar temperature and $\log g$, making them useful age discriminators for identifying young members of moving groups.  When combined with an estimate of $[$Fe/H$]$, the equivalent width of the \naone doublet also yields an excellent estimate of an M star's absolute $K$-band magnitude, and may have some sensitivity to $\alpha$ enhancement \citep{terrien15}.  With regards to activity, \citet{kafka06} showed that the equivalent widths of the \naone doublet were anticorrelated with \ha linewidths for M dwarfs in Praesepe, suggesting at least some correlation of the doublet with a star's mean activity level.  Thus, the sodium doublet merits careful consideration as a primary activity indicator in the NIR.

Since the two lines of the \naone doublet are produced by the same atomic transition, and are equally reliable in terms of visibility, telluric contamination, etc., we have created a combined index, \wnap, which is simply the average of the $W$ index for each individual line.  We did consider each line individually, and found no differences between their individual sensitivities to activity.  We note also that the UVES spectra of Proxima have a gap in wavelength coverage from approximately 8212-8365\AA, which restricts our choice of a redward reference band for \wnap.

\begin{center}\emph{The \kone NIR Doublet}\end{center}

As with the sodium doublet, \citet{kafka06} found that the equivalent widths of the neutral potassium doublet at $\lambda\lambda = 7664.90, 7698.96$\AA~were anticorrelated with \ha for M stars in Praesepe, indicating activity sensitivity.  Like the \naone lines, the \kone doublet is a strong spectral feature for all M subtypes, making for easy identification and measurement.  The primary drawback to using the \kone lines is that they are contaminated with telluric O$_2$ lines, which are significantly more difficult to correct than other common species (see Section \ref{sec:telluric}).

Since each of the \kone lines show qualitatively identical responsiveness to activity for Proxima, we have again created a combined index \wkp, which is the average $W$ value for each of the lines.

\begin{center}\emph{The Paschen Delta Line}\end{center}

At $\lambda = 10049.4$\AA, P$\delta$ is the reddest of our candidate tracers, and lies near the limit of the UVES spectra's coverage range.  P$\delta$ has been shown to trace flare activity on M stars \citep[e.g.][]{schmidt12}, and we therefore sought to determine whether it might also be sensitive to the low-level activity variability seen for Proxima and other M stars in periods of relative quiescence.

\section{\bf Telluric Correction} \label{sec:telluric}

The near-infrared sky is considerably more contaminated by telluric absorption than the optical.  In particular, the wavelength regions containing the \kone and \naone NIR doublets contain strong absorption from the Earth's atmosphere.  Thus, in order to make reliable measurements of \wk and \wnap, it was necessary to first correct the UVES spectra for telluric absorption.

Telluric correction was performed using TERRASPEC \citep{bender12,lockwood14}.  TERRASPEC uses the LBLRTM radiative transfer code \citep{clough05} to generate a synthetic telluric absorption function that is customized for an observer's altitude, the target zenith angle, and the telluric conditions integrated over a given science observation.  LBLRTM uses the HITRAN 2008 database of molecular lines \citep{rothman09} and can access a variety of standard atmospheric models.  The telluric absorption function is combined in a forward model with the spectrograph instrument profile (IP), parameterized as a series of overlapping Gaussians \citep[e.g.][]{valenti95,endl00}.  TERRASPEC uses a non-linear least-squares minimization to optimize the integrated column depths of the absorbers and IP parameters to best match the observed science spectrum.  This process facilitates a high-quality telluric correction that can be computed directly from the science spectrum, and does not rely on observed `telluric standard' spectra being available. To process the UVES spectra, we used the standard mid-latitude atmospheric model and an IP parameterized as a central Gaussian surrounded on either side by two satellite Gaussians.

In order to be computationally efficient, we have only performed telluric correction on sections of the UVES spectra containing the \kone and \naone doublets and their reference bands.  In Figure \ref{fig:terraspec}, we show the results of the TERRASPEC correction for an example spectrum around the \kone lines.  Also included in Figure \ref{fig:terraspec} is the model to the IP, which appears typical of profiles recovered for high-resolution spectrographs \citep[e.g.][]{endl00}.  For regions containing the other lines we have considered, the telluric lines are very weak, so we do not expect significant contamination from telluric lines in these regions.

The spectral region near the NIR \naone lines is dominated by lines from atmospheric water, the relative strengths of which vary considerably across the three nights of intensive observation.  Around the \kone doublet, the spectra are littered with lines from O$_2$, which are more difficult to correct, since their profiles diverge markedly from a Voigt profile.  This may ultimately prove to be a limiting factor for using the \kone doublet as an activity tracer, although we note that we find little difference between our \wk values before and after telluric correction.

\section{\bf Notes on Proxima's Activity} \label{sec:activity}

The primary goal of this paper is not to discuss stellar activity on Proxima in detail.  However, since our use of Proxima as a testbed for new activity tracers relies on at least some stellar variability, we will make some brief remarks regarding Proxima's activity.

Even over short time scales, the spectral activity indicators show significant variability.  The second night of intensive UVES observations--the quietest of the three nights--shows an RMS scatter in \iha of 0.017, with an average measurement uncertainty of just 0.004.  High-cadence HARPS observations from the Cool Tiny Beats survey are similarly variable.  As mentioned previously, this short-term variability is advantageous for our study, as it provides a broad range of activity levels across which to evaluate candidate activity tracers.

The most problematic manifestation of stellar activity for contemporary Doppler searches for exoplanets around M stars is the periodicity induced by rotating starspots and active regions.  The relatively sparse sampling of Proxima by the HARPS and UVES RV surveys, coupled with the high short-term variability of the spectral activity indices, makes determination of the stellar rotation period with absorption-line indices difficult.  Fortunately, there are 1049 high-quality $V$-band photometric observations of Proxima taken over 9 years as part of the All Sky Automated Survey (ASAS) currently available in the ASAS public archive \citep{pojmanski97}.  These photometric data clearly reveal the rotation period at 84 days, consistent with the $82.5$-day period derived by \citet{kiraga07} from the first 609 ASAS observations, and with the $83.2$-day period found by \citet{suarez16} using the full ASAS data set.

\begin{figure}
\begin{center}
\includegraphics[width=1.0\columnwidth]{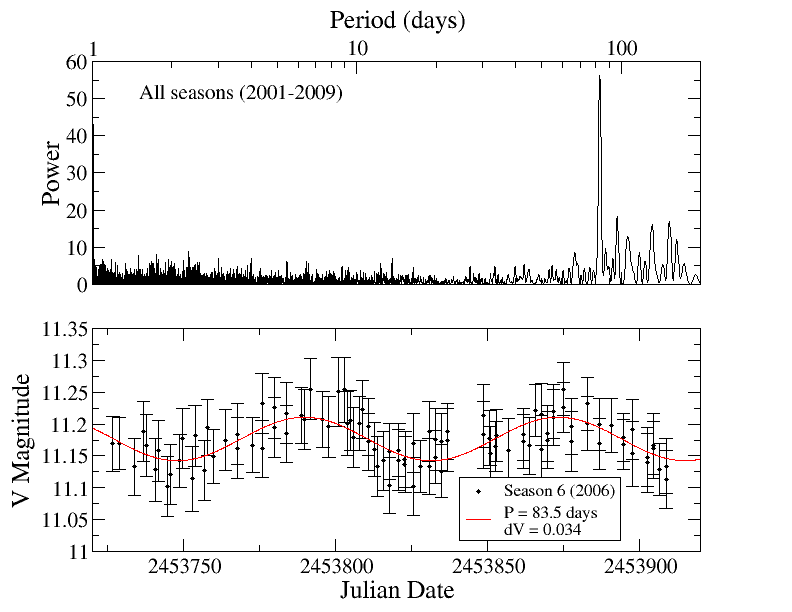}
\caption{\label{fig:rotation}
\footnotesize \emph{Top}: Generalized Lomb-Scargle periodogram of $V$-band ASAS photometry of Proxima.  The most significant peak is the 84-day rotation period.  \emph{Bottom}: ASAS photometry from the 2006 observing season.  We include a sinusoidal model to this season alone as a solid red curve.}
\end{center}
\end{figure}

In Figure \ref{fig:rotation}, we show the generalized Lomb-Scargle periodogram \citep{zk09} of all 9 seasons of ASAS photometry.  The Lomb-Scargle periodogram is an adaptation of the Fourier transform that excels at identifying periodicity in irregularly-sampled time series.  The 84-day periodicity is clearly distinguished from all other candidate periods.  In the same Figure, we highlight the photometric rotation signal from the 2006 observing season, where the signal is especially well resolved, with an amplitude of $0.03$ magnitudes.

\begin{figure}
\begin{center}
\includegraphics[width=1.0\columnwidth]{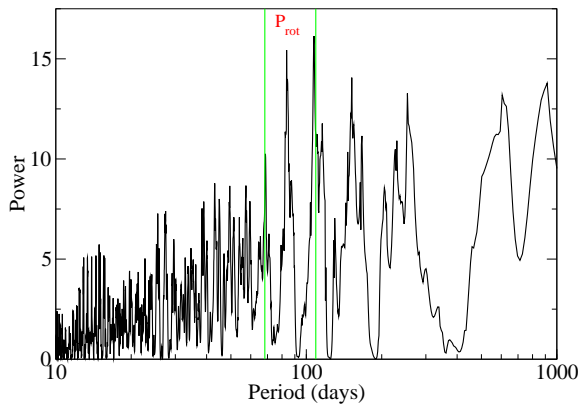}
\caption{\label{fig:fwhm}
\footnotesize Generalized Lomb-Scargle periodogram of the FWHM values from the HARPS spectra prior to the Cool Tiny Beats observations.  The most significant peaks occur at the 84-day rotation period and one of its yearly aliases, which are marked with green lines.  The appearance of the rotation period in FWHM strongly suggests the stellar rotation signal will be present in RV observations.}
\end{center}
\end{figure}

We note that \citet{suarez15} find a candidate rotation period of 117 days using the \catwo H\&K and \ha lines in the HARPS spectra.  We agree that this period appears at high significance for the HARPS \shk and \iha values prior to the addition of the high-cadence observations from the Cool Tiny Beats survey.  The short-term variability seen in the Cool Tiny Beats observations dominates the power spectrum of the combined data set, making the 117-day period more difficult to identify.  However, the photometric observations greatly outnumber those from HARPS, and have considerably better phase coverage.  We analyzed the ASAS observations season-by-season in case the true period is actually 117 days, and the ASAS periodogram prefers 84 days due to phase incoherence of the rotation signal.  Instead, for every season in which we recovered a statistically significant signal, the preferred period was at or near 84 days, thus confirming the shorter rotation period.  We suspect the 117-day signal may be caused by incomplete sampling of the 84-day stellar rotation, which is more difficult to detect with spectral indicators because of the short-term variability.  This hypothesis is consistent with the findings of \citet{collins16}, who explore the origin of the 117-day periodicity in more detail.

We find evidence that--as is common for most targets of high-precision RV surveys--the stellar rotation signal will be present in RVs of Proxima.  In addition to absorption-line indices, we have examined the full-width at half maximum (FWHM) of the cross-correlation function for the HARPS spectra.  The FWHM, which is produced by the standard HARPS data reduction pipeline and provided in the image headers, can be interpreted as the mean stellar line width for the observation.  If activity-dependent variability alters the stellar line shapes, the FWHM should trace these changes.  Since changes in stellar line widths must be unreasonably symmetric to avoid altering the measured RV, any periodic signal observed in FWHM will likely appear in RV as well.  In Figure \ref{fig:fwhm}, we show the generalized Lomb-Scargle periodogram of the HARPS FWHM values taken prior to the Cool Tiny Beats survey; again, we have excluded the high-cadence observations because they cause short-term variability to dominate the power spectrum.  The most significant peaks occur near the 84-day rotation period and its yearly alias at 109 days.  FWHM variations are difficult to measure for very small activity-induced RV shifts, so this detection suggests at least a moderate effect.  Thus, the periodicity observed in FWHM suggests that the RVs of Proxima should be modulated by the stellar rotation.  Indeed, the RVs measured by the Pale Red Dot program described in \citet{ae16} show a trend in addition to the 11.2-day planet signal, which appears to correlate with quasi-simultaneous photometry (specifically the $F^\prime$ statistic, which approximates the derivative of the photometric light curve).  We speculate that stellar rotation may be the origin of this trend, but this suggestion requires additional analysis that is outside the scope of this study.  Importantly, though, none of the stellar periodicities observed here or elsewhere in the literature \citep[e.g.][]{suarez16,collins16,ae16} occur near the 11.2-day period of Proxima b, greatly reducing the likelihood that the proposed planet is a false positive induced by stellar activity.

\begin{figure}
\begin{center}
\includegraphics[width=1.0\columnwidth]{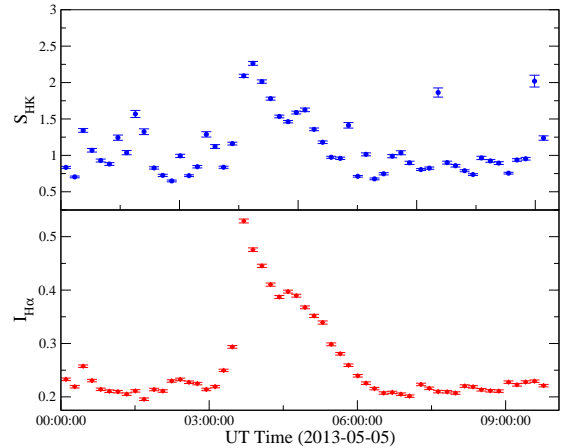}
\caption{\label{fig:harpsflare}
\footnotesize A large flare on Proxima, as observed by HARPS in \shk (\emph{top}) and \iha (\emph{bottom}).}
\end{center}
\end{figure}

\citet{cincunegui07a} found evidence for a 442-day magnetic cycle for Proxima, based on 60 observations of the \catwo H\&K lines taken between 1999 and 2006 with the REOSC Spectrograph.  We are unable to confirm this cycle with the data considered herein.  However, we note that there is little overlap between the REOSC and HARPS/UVES observations, so it is possible that Proxima entered a quiet phase shortly after the REOSC observations concluded, making the magnetic cycle difficult to detect.  \citet{suarez16} observe a 6.8-year periodicity in the ASAS light curve, which they also attribute to a magnetic cycle.  We agree that the ASAS photometry shows a periodicity on this timescale, but again see no evidence of a corresponding spectroscopic signal.

One final issue to consider with regards to Proxima's activity is its frequent flares.  \citet{davenport16} observe Proxima to exhibit small flares 63 times per day, with larger flares occurring many times per year.  Proxima's frequent flaring presents a challenge for the detection--or confident non-detection--of a transit for Proxima b.  The high-cadence observations from both HARPS and UVES show a number of flares, and each set of spectra has one night that is mostly dominated by a large flare.  The large flare from the third night of UVES observations is thoroughly described in \citet{fuhrmeister11}, but we show the large flare from the HARPS data as traced by \ha and \catwo H\&K in Figure \ref{fig:harpsflare}.  The difference in appearance for the flare in the calcium and \ha lines is visually striking; we discuss the varying response of the atomic lines to flares further in Section \ref{sec:flare}.

\section{\bf Evaluation of Potential Activity Indicators} \label{sec:tracers}

\subsection{Equivalence of \ha and \shk}

The nonstandard instrument configuration used by \citet{fuhrmeister11} for their UVES observations is advantageous in that the resultant spectra include \ha as well as all of our candidate activity tracer lines.  This allows us to compare each line's responsiveness to changes in the stellar activity level, using \iha as the known standard with which to measure activity.  However, it is important we establish that \iha is itself suitably sensitive to magnetic activity for Proxima.  While often an excellent activity metric, \iha has been observed to be uncorrelated with \shk for some M stars \citep[e.g.][]{cincunegui07b}.

\begin{figure}[h!]
\begin{center}
\includegraphics[width=1.0\columnwidth]{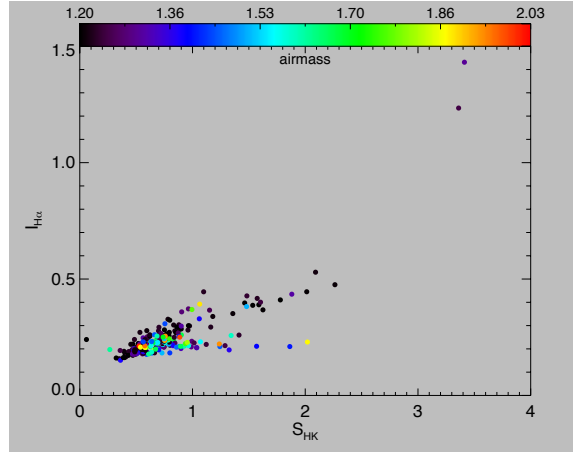}
\caption{\label{fig:shk_iha}
\footnotesize \iha as a function of \shk for the 261 HARPS spectra.  Points are color-coded by the airmass at the start of each observation, and error bars are excluded for visual clarity.}
\end{center}
\end{figure}

The optical HARPS spectra offer an ideal avenue through which to check the equivalence (or lack thereof) of \iha and the other optical activity tracers, as they acquire \catwo H\&K, \naone D, and \ha simultaneously.  In Figure \ref{fig:shk_iha}, we show \iha as a function of \shk for the 261 HARPS spectra.  The quantities appear to be clearly correlated, but some high-airmass observations deviate from this correlation.  This is most likely due to the fact that at high airmass and in poor seeing, the signal to noise in the continuum near the H\&K lines is even lower than normal, making the \shk measurement unreliable.  When restricting the set to low-airmass observations, the correlation of \iha with \shk is clear.  This result is consistent with the results of \citet{cincunegui07a}, who also found the fluxes in the \ha and H\&K lines to be correlated for Proxima.

\begin{figure}[h!]
\begin{center}
\includegraphics[width=1.0\columnwidth]{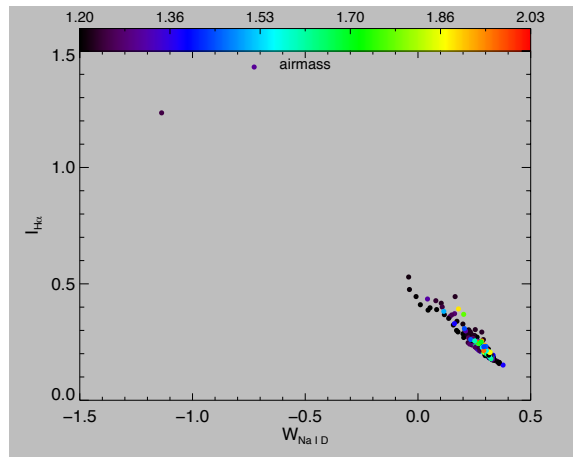}
\caption{\label{fig:id_iha}
\footnotesize \iha as a function of \id for the 261 HARPS spectra.  Points are color-coded by the airmass at the start of each observation, and error bars are excluded for visual clarity.}
\end{center}
\end{figure}

Comparison with the optical sodium doublet provides additional confirmation that \iha is representative of the optical activity tracers.  As shown in Figure \ref{fig:id_iha}, \iha is very tightly correlated with \wnad over all observations.  The adherence of the high-airmass points to this relationship also offers further evidence that the scatter in the \ihap-\shk relationship is largely due to systematic errors in \shk at high airmass.

Thus, we are satisfied that the \ha line is functionally equivalent to \shk or \wnad for measuring changes in stellar magnetic activity on Proxima.  This allows us to ``bootstrap" our comparison to the NIR absorption lines, using \iha as a proxy for any optical tracer.

\subsection{Exclusion of Flare Observations}
\label{sec:flare}

\begin{figure}[h!]
\begin{center}
\includegraphics[width=1.0\columnwidth]{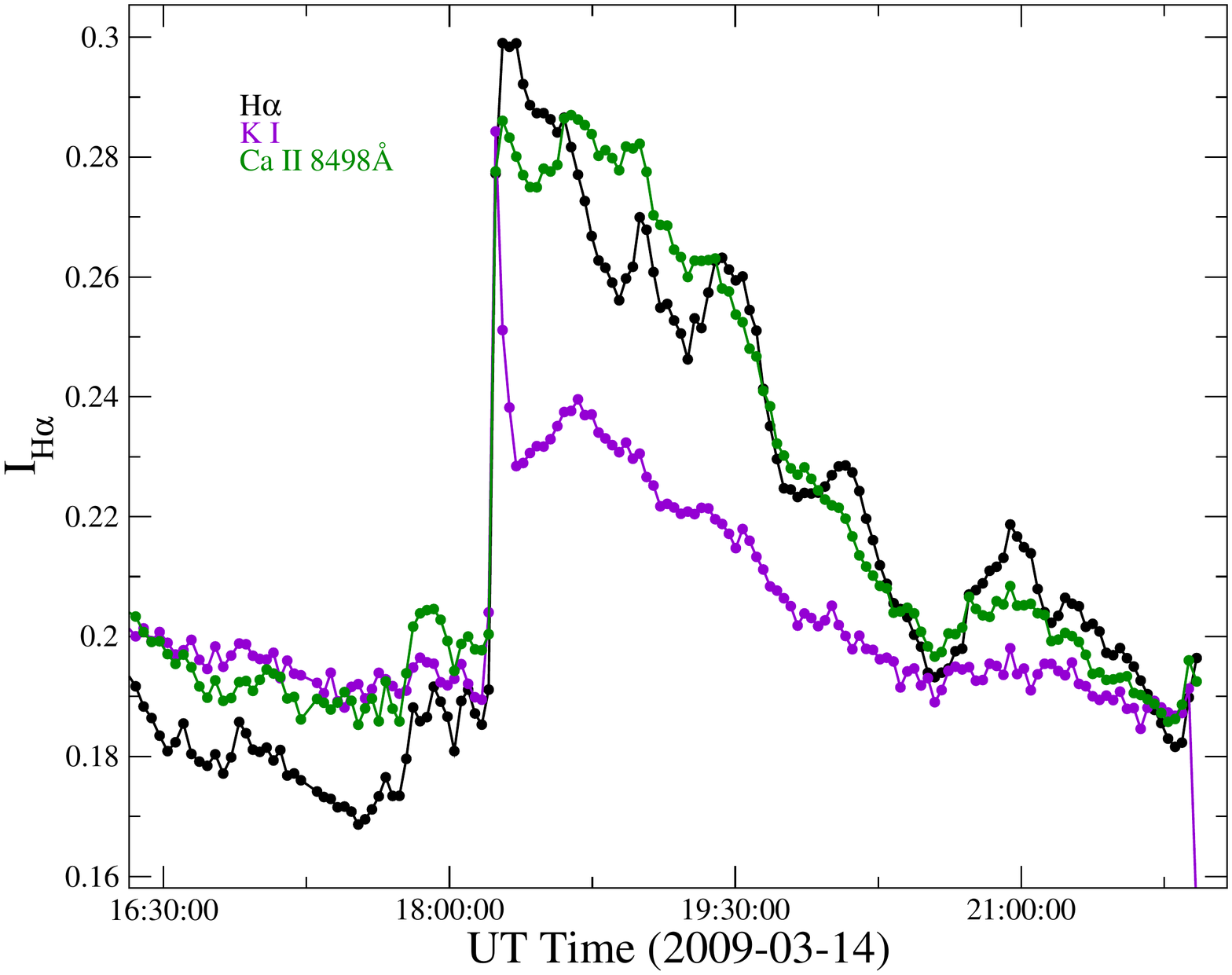}
\caption{\label{fig:flare}
\footnotesize The flare on night 3 of the high-cadence UVES observations, as traced by \ihap, \wkp, and \wcaonep.  Values of \wk and \wcaone are scaled and inverted to facilitate visual comparison with \ihap.}
\end{center}
\end{figure}

A preliminary examination of the candidate NIR activity tracers shows clear correlations of \wk and the three calcium lines with \ihap.  However, the correlations are markedly different for the third night than for the first two. As detailed in \citet{fuhrmeister11}, the majority of spectra from this third night were taken during the onset and decay of a large stellar flare.  Each of the lines evaluated herein responds to the flare differently; in Figure \ref{fig:flare} we show the flare as observed by \ihap, \wkp, and \wcaonep.  Such disparate morphologies are typical of M dwarf flares, especially during the decay phase; the fluxes of chromospheric lines are frequently observed to decrease at different rates after the initial rise, and may not even peak at the same time \citep[e.g.][and references therein]{hawley91,kowalski13}.

Because we are primarily interested in how well the candidate tracers respond to the low-level magnetic variability that causes the bulk of the astrophysical noise in RV measurements, and because there are well-understood physical reasons why chromospheric lines of different species should behave differently during flares, we have excluded all spectra from this third night from our analysis in the next section.  Removing the flare observations eliminates the dominant source of scatter in the observed relations for the \kone and \catwo lines with \hap.

Although we have removed the flare observations from this analysis, we note that the overall morphology of the correlations of our candidate lines with \iha does not change even during the flare.  Rather, the correlations are simply offset from the more quiescent relationship, and have more scatter.  Thus, RV surveys relying on the NIR activity tracers presented herein will be as sensitive to both flares and low-level variability as those using the canonical optical indices.

While it is certainly valuable to know if a star is flaring during an observation, it appears that Doppler measurements of M stars may not be significantly affected by flares.  \citet{ae16} find that RVs measured from the Cool Tiny Beats observations of Proxima in May 2013 were not altered by the flare (shown here in Figure \ref{fig:harpsflare}) that occurred during those exposures.  We speculate there are two predominant reasons why the RVs appear to be immune to flares.  First, it is common for deep chromospheric lines--which are most sensitive to flares--to be omitted from masks or line lists for Doppler pipeline codes.  Also, M dwarf flares produce additional flux mostly at bluer wavelengths, where those stars are intrinsically faint \citep{kowalski13}.  Thus, the flare contribution to the spectral orders used to compute RVs for M stars is likely minimal.

\subsection{Behavior of Candidate Tracers}

In restricting our analysis to the first two nights of high-cadence UVES observations, we are left with 383 spectra spanning a range of $0.136$ in \ihap.  For comparison, the peak-to-peak \iha amplitude of the rotation signal observed for Kapteyn's star--another low-mass nearby M star--is just $0.006$~\citep{robertson15}.  Thus, while our analysis is limited to just two nights' worth of data, it covers a greater range of activity levels than will ever be observed for the most ideal targets for discovering exoplanets around M stars with RV.

\begin{figure*}
\begin{center}
\includegraphics[width=1.8\columnwidth]{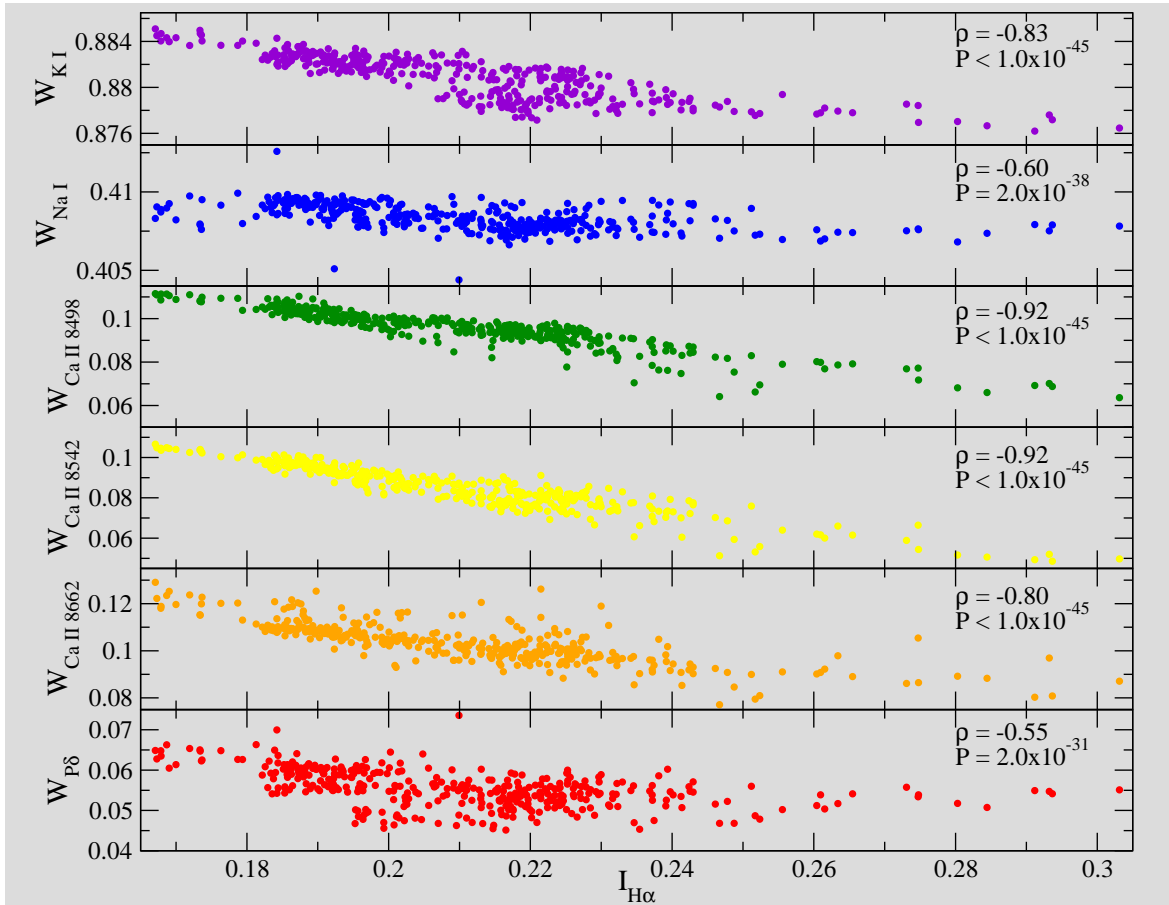}
\caption{\label{fig:tracer_comp}
\footnotesize Measurements of our candidate activity indicators as a function of \ihap.  For each indicator we give the Spearman rank correlation coefficient $\rho$.  \wk and the three calcium indices are all closely anticorrelated with \ihap.}
\end{center}
\end{figure*}

We show each of our 6 candidate tracers as a function of \iha in Figure \ref{fig:tracer_comp}.  Note again that because our $W$ index is essentially an equivalent width, it decreases as the flux in the line increases.  Hence, the $W$ indices appear anticorrelated with \ihap, which increases as the flux in the \ha line increases.

For each relation between a candidate tracer $W$ and \iha we have included the value of the Spearman rank correlation coefficient $\rho$.  The Spearman coefficient, which determines how closely a pair of ranked variables obey a monotonic function, ranges between 0 (not monotonic) and $\pm 1$~(perfectly monotonic).  We chose the Spearman statistic because it does not assume that the data are normally distributed or follow a linear relationship, as do other probabilistic tests of correlation such as the Pearson correlation coefficient.  While we find that the observed relations are well described by a linear fit, we do not wish to require this assumption.  Along with each value of $\rho$ we include an estimate of $P$, the probability that we would derive the observed value of $\rho$ for 383 observations of variables which are in fact uncorrelated.  In some cases, $P$ falls below machine precision, and is therefore limited to $P < 10^{-45}$.

The $\rho$ values of all 6 indicators suggests significant anticorrelation with \ihap.  However, the purpose of this investigation is not to determine whether the lines are sensitive to activity; indeed, as discussed in Section \ref{sec:candidates}, all of the lines considered are known to have at least some sensitivity to activity.  Rather, our goal is to find one or more tracers that are essentially equivalent to the well-established optical indicators \shk and \ihap.  To that end, \wk and all three of the calcium indicators appear to be excellent options, with the two bluest calcium lines in particular following almost perfectly monotonic ($\rho < -0.9$) relations with \ihap.

In addition to perhaps being inherently less sensitive to activity than \wcaone and \wcatwop, there are some well-understood sources of additional noise in our measurements of \wk and \wcathree that lead to a somewhat less perfect correlation with \ihap.  As shown in Figure \ref{fig:lines}, the \catwo 8662\AA~line is heavily blended with the \ion{Fe}{1} line at 8661.9\AA, which restricts the size of the window we can use to sum the flux without introducing contamination.  For \wkp, it is likely that imperfect telluric correction is contributing noise to the measurements.  In addition, for the \kone lines we found that for Proxima, window sizes smaller than 1\AA~produced tighter relations with \ihap.  However, we have noticed that for later-subtype M stars, the \kone lines tend to have broad central reversals, necessitating the 1\AA~windows.  Thus, in an effort to make our results as broadly applicable as possible, we have adopted the 1\AA~windows for Proxima.  While in the case of Proxima \wcaone and \wcatwo are clearly the most ideal activity indicators, there are likely scenarios in which \wk or \wcathree will prove more suitable.  We discuss one such example in Section \ref{sec:trappist1} below.

\subsection{Application for late-M stars: the example of TRAPPIST-1} \label{sec:trappist1}

The goal of this investigation is to provide activity indicators that will be useful to Doppler surveys targeting mid-late M stars in the near infrared.  While we have shown that \wk and the calcium indices are excellent choices for Proxima, a mid-M dwarf, full confirmation of their suitability for later subtypes must await the availability of significant numbers of observations on late-M stars with new NIR spectrographs.  Briefly, though, we highlight spectroscopy of the M8 planet host TRAPPIST-1 as preliminary evidence that \wk may be especially useful for late-M dwarfs.

TRAPPIST-1, alternatively designated 2MASS J23062928-0502285, was recently shown by \citet{gillon16} to host a system of three approximately Earth-sized transiting exoplanets.  Such a system represents the exemplary science case for NIR Doppler spectrographs.  As part of their NIR RV survey with UVES, \citet{barnes14} acquired 4 spectra of TRAPPIST-1 in July 2012 using an instrumental configuration similar to that used by \citet{fuhrmeister11} for Proxima.  This therefore provides an opportunity to again compare the NIR activity tracers to \ihap.

The observations of late-M stars from \citet{barnes14} achieve a significantly lower S/N than the spectra of Proxima examined herein.  The lower S/N has two significant consequences for our evaluation of activity indicators.  First, the S/N near the \catwo NIR triplet is particularly low, making index measurements of these lines unreliable.  Second, the lower S/N in the deep line cores (including near the \kone cores) makes the correction of telluric absorption by O$_2$ with TERRASPEC especially difficult.  For these reasons, we do not consider all of the late-M stars from \citet{barnes14} in this work.  However, because we were able to achieve a reasonable telluric correction in the \kone region for TRAPPIST-1, and because of the increased scientific value of this star, we will briefly remark on the \kone lines of TRAPPIST-1.

As mentioned previously, the \kone lines for TRAPPIST-1 contain broad central reversals, necessitating a 1\AA~window for the measurement of \wkp.  Since a cursory inspection of the other late-M spectra from \citet{barnes14} shows this to be typical of the later M types, we have made the 1\AA~window the default width for our \wk index.

\begin{figure}
\begin{center}
\includegraphics[width=1.0\columnwidth]{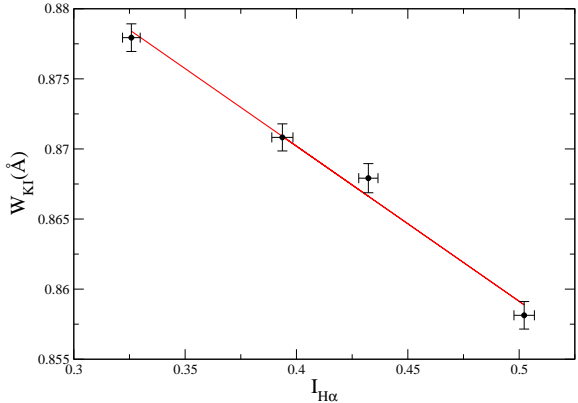}
\caption{\label{fig:trappist1}
\footnotesize \wk as a function of \iha for the planet-host M8 dwarf TRAPPIST-1.  A linear fit to the data is shown as a solid red line.  The observed anticorrelation tentatively suggests that \wk is sensitive to activity variations even in late M stars.}
\end{center}
\end{figure}

Figure \ref{fig:trappist1} shows \wk as a function of \iha for the four UVES spectra of TRAPPIST-1.  As for Proxima, the quantities appear to be anticorrelated.  However, with just 4 observations, the anticorrelation is much more difficult to establish statistically.  A Pearson correlation test yields a correlation coefficient $r = -0.99$ which, for a sample size $N = 4$ translates to a $P$-value of 0.01.  While the small sample size limits our results to only $\sim3\sigma$ significance, it is nonetheless encouraging to see that \wk appears to be a sensitive activity indicator even at very late M types.

\section{\bf Discussion} \label{sec:discussion}

The strong anticorrelation of the NIR calcium indices with \iha is not in itself surprising.  The \catwo triplet has been shown previously to be sensitive to activity, including for this same UVES data set by \citet{barnes14}.  What is remarkable, though, is the degree of sensitivity and the ease of measurement, particularly for the two bluest lines.  Barnes et al.~removed the mean spectrum from each observation before measuring the equivalent widths of the lines, but we find this is unnecessary.  Despite the fact that the $8498$\AA~and $8542$\AA~lines are frequently completely invisible against the pseudocontinuum, simply summing the flux within windows centered on their wavelengths provides an activity indicator that is essentially equivalent to \ihap.  Furthermore, while Barnes et al.~concluded that the \catwo triplet was useful for monitoring flare activity, we see that its applicability extends far beyond just detecting flares; virtually every small activity feature observed in \iha is matched exactly by the calcium lines.  Activity signals at these lower amplitudes are the ones most likely to confound Doppler exoplanet searches, especially since their photometric counterparts may be absent or difficult to detect.  Thus, we anticipate the NIR calcium triplet to be of immense value to the Doppler community.

The NIR \kone doublet will not be within the wavelength range of several of the upcoming NIR Doppler spectrographs, including HPF and SPIRou, but its activity sensitivity is nonetheless interesting for a number of reasons.  Some upcoming instruments, such as MAROON-X \citep{seifahrt16} and \emph{Veloce}\footnote{http://newt.phys.unsw.edu.au/$\sim$cgt/Veloce/Veloce.html}, are optimized for red-optical wavelength coverage in order to target early M stars as a primary science objective.  Other spectrographs, such as NEID and ESPRESSO, will primarily target solar-type stars, but their wavelength coverage will extend redward of the potassium doublet.  These instruments will certainly observe at least some early-mid M stars, and the \kone feature may prove to be a useful diagnostic, especially when--as we observe for TRAPPIST-1--the S/N near the NIR calcium triplet is too low to make useful measurements of those lines.  Additionally, while the \kone doublet at $\sim7700$\AA~is inaccessible to several dedicated NIR spectrometers, its sensitivity to activity raises the intriguing possibility that the four strong \kone lines near $\lambda = 12000$\AA~are similarly sensitive.  If so, they are optimally placed in wavelength space for NIR Doppler instruments, and are strong features for all mid-late M subtypes \citep[see, e.g.,][]{deshpande12}, and should be an ideal activity tracer for mid-late M stars in the NIR.

It is important to emphasize that although the results shown herein are encouraging for upcoming NIR Doppler surveys, they must be considered preliminary pending analysis of the data those surveys will produce.  With CARMENES now conducting its science campaign, and HPF scheduled to be on-sky within the year, we will soon have time-resolved NIR observations of mid-late M stars over baselines much longer than the observations of Proxima examined in this work.  Our proposed \kone and \catwo indices must be validated over all M subtypes, and for multiple magnetic phenomena (starspots, activity cycles, etc.).  Observations with a true NIR spectrograph--as opposed to with the red-optical coverage of UVES--will also provide an opportunity to evaluate even redder activity-sensitive lines.  The redder \kone lines ($\lambda\lambda = 11690, 11771, 12435, 12522$ \AA) mentioned above should all be examined, as well as the \ion{He}{1} line at $\lambda = 10830$\AA, and the stronger NIR hydrogen lines such as P$\beta$ and P$\gamma$ \citep[e.g.][]{schmidt12}.  In the meantime, the indices we propose here should prove valuable in the early stages of these instruments' Doppler surveys as the best known spectral indicators of activity, thus offering a means of discerning potential exoplanet signals from activity-induced false positives.

\section{\bf Conclusions} \label{sec:conclusions}

We have used high-cadence optical and near-infrared spectroscopy of Proxima Centauri to examine the sensitivity of atomic absorption features to stellar magnetic activity.  By using \ha as a known proxy for stellar activity, we found that the \catwo lines at $\lambda\lambda = 8498, 8542$\AA~and the \kone doublet near $\lambda = 7700$\AA~very closely match the stellar variability seen in \hap.  The \kone doublet is also tightly correlated with \ha for four UVES observations of the M8 planet-host star TRAPPIST-1, suggesting that feature may be sensitive to activity for even the latest M stars.  The indices we define for these lines are thus excellent candidates for preliminary activity metrics in NIR and red-optical Doppler surveys.

\begin{acknowledgements}
We thank the anonymous referee for his or her insightful comments.  This work was performed in part under contract with the California Institute of Technology (Caltech)/Jet Propulsion Laboratory (JPL) funded by NASA through the Sagan Fellowship Program executed by the NASA Exoplanet Science Institute.  CFB acknowledges partial support from NSF grant AST-1517592.
  This work was partially supported by funding from the Center for Exoplanets and Habitable Worlds. The Center for Exoplanets and Habitable Worlds is supported by the Pennsylvania State University, the Eberly College of Science, and the Pennsylvania Space Grant Consortium.  We acknowledge support from NSF grants AST 1006676, AST 1126413, AST 1310885, and the NASA Astrobiology Institute (NNA09DA76A) in our pursuit of precision radial velocities in the NIR.  This work was performed in part at the Aspen Center for Physics, which is supported by National Science Foundation grant PHY-1066293.
\end{acknowledgements}

\bibliography{proxima_nir.bib}

\begin{thebibliography}{}
\expandafter\ifx\csname natexlab\endcsname\relax\def\natexlab#1{#1}\fi

\bibitem[{{Andretta} {et~al.}(2005){Andretta}, {Bus{\`a}}, {Gomez}, \&
  {Terranegra}}]{andretta05}
{Andretta}, V., {Bus{\`a}}, I., {Gomez}, M.~T., \& {Terranegra}, L. 2005, \aap,
  430, 669

\bibitem[{{Andretta} {et~al.}(1997){Andretta}, {Doyle}, \&
  {Byrne}}]{andretta97}
{Andretta}, V., {Doyle}, J.~G., \& {Byrne}, P.~B. 1997, \aap, 322, 266

\bibitem[{{Anglada-Escud{\'e}} {et~al.}(2014){Anglada-Escud{\'e}}, {Arriagada},
  {Tuomi}, {Zechmeister}, {Jenkins}, {Ofir}, {Dreizler}, {Gerlach}, {Marvin},
  {Reiners}, {Jeffers}, {Butler}, {Vogt}, {Amado},
  {Rodr{\'{\i}}guez-L{\'o}pez}, {Berdi{\~n}as}, {Morin}, {Crane}, {Shectman},
  {Thompson}, {D{\'{\i}}az}, {Rivera}, {Sarmiento}, \& {Jones}}]{ae14}
{Anglada-Escud{\'e}}, G., {Arriagada}, P., {Tuomi}, M., {et~al.} 2014, \mnras,
  443, L89

\bibitem[{{Anglada-Escud{\'e}} {et~al.}(2016){Anglada-Escud{\'e}}, {Amado},
  {Barnes}, {Berdi{\~n}as}, {Butler}, {Coleman}, {de La Cueva}, {Dreizler},
  {Endl}, {Giesers}, {Jeffers}, {Jenkins}, {Jones}, {Kiraga}, {K{\"u}rster},
  {L{\'o}pez-Gonz{\'a}lez}, {Marvin}, {Morales}, {Morin}, {Nelson}, {Ortiz},
  {Ofir}, {Paardekooper}, {Reiners}, {Rodr{\'{\i}}guez},
  {Rodr{\#943}guez-L{\'o}pez}, {Sarmiento}, {Strachan}, {Tsapras}, {Tuomi}, \&
  {Zechmeister}}]{ae16}
{Anglada-Escud{\'e}}, G., {Amado}, P.~J., {Barnes}, J., {et~al.} 2016, \nat,
  536, 437

\bibitem[{{Artigau} {et~al.}(2014){Artigau}, {Kouach}, {Donati}, {Doyon},
  {Delfosse}, {Baratchart}, {Lacombe}, {Moutou}, {Rabou}, {Par{\`e}s},
  {Micheau}, {Thibault}, {Reshetov}, {Dubois}, {Hernandez}, {Vall{\'e}e},
  {Wang}, {Dolon}, {Pepe}, {Bouchy}, {Striebig}, {H{\'e}nault}, {Loop},
  {Saddlemyer}, {Barrick}, {Vermeulen}, {Dupieux}, {H{\'e}brard}, {Boisse},
  {Martioli}, {Alencar}, {do Nascimento}, \& {Figueira}}]{artigau14}
{Artigau}, {\'E}., {Kouach}, D., {Donati}, J.-F., {et~al.} 2014, in \procspie,
  Vol. 9147, Ground-based and Airborne Instrumentation for Astronomy V, 914715

\bibitem[{{Baliunas} {et~al.}(1995){Baliunas}, {Donahue}, {Soon}, {Horne},
  {Frazer}, {Woodard-Eklund}, {Bradford}, {Rao}, {Wilson}, {Zhang}, {Bennett},
  {Briggs}, {Carroll}, {Duncan}, {Figueroa}, {Lanning}, {Misch}, {Mueller},
  {Noyes}, {Poppe}, {Porter}, {Robinson}, {Russell}, {Shelton}, {Soyumer},
  {Vaughan}, \& {Whitney}}]{baliunas95}
{Baliunas}, S.~L., {Donahue}, R.~A., {Soon}, W.~H., {et~al.} 1995, \apj, 438,
  269

\bibitem[{{Barnes} {et~al.}(2012){Barnes}, {Jenkins}, {Jones}, {Rojo},
  {Arriagada}, {Jord{\'a}n}, {Minniti}, {Tuomi}, {Jeffers}, \&
  {Pinfield}}]{barnes12}
{Barnes}, J.~R., {Jenkins}, J.~S., {Jones}, H.~R.~A., {et~al.} 2012, \mnras,
  424, 591

\bibitem[{{Barnes} {et~al.}(2014){Barnes}, {Jenkins}, {Jones}, {Jeffers},
  {Rojo}, {Arriagada}, {Jord{\'a}n}, {Minniti}, {Tuomi}, {Pinfield}, \&
  {Anglada-Escud{\'e}}}]{barnes14}
---. 2014, \mnras, 439, 3094

\bibitem[{{Bastien} {et~al.}(2014){Bastien}, {Stassun}, {Pepper}, {Wright},
  {Aigrain}, {Basri}, {Johnson}, {Howard}, \& {Walkowicz}}]{bastien14}
{Bastien}, F.~A., {Stassun}, K.~G., {Pepper}, J., {et~al.} 2014, \aj, 147, 29

\bibitem[{{Bean} {et~al.}(2010){Bean}, {Seifahrt}, {Hartman}, {Nilsson},
  {Wiedemann}, {Reiners}, {Dreizler}, \& {Henry}}]{bean10}
{Bean}, J.~L., {Seifahrt}, A., {Hartman}, H., {et~al.} 2010, \apj, 713, 410

\bibitem[{{Bender} {et~al.}(2012){Bender}, {Mahadevan}, {Deshpande}, {Wright},
  {Roy}, {Terrien}, {Sigurdsson}, {Ramsey}, {Schneider}, \&
  {Fleming}}]{bender12}
{Bender}, C.~F., {Mahadevan}, S., {Deshpande}, R., {et~al.} 2012, \apjl, 751,
  L31

\bibitem[{{Blake} {et~al.}(2015){Blake}, {Johnson}, {Plavchan}, {Sliski},
  {Wittenmyer}, {Eastman}, \& {Barnes}}]{blake15}
{Blake}, C., {Johnson}, J., {Plavchan}, P., {et~al.} 2015, in American
  Astronomical Society Meeting Abstracts, Vol. 225, American Astronomical
  Society Meeting Abstracts, 257.32

\bibitem[{{Boisse} {et~al.}(2009){Boisse}, {Moutou}, {Vidal-Madjar}, {Bouchy},
  {Pont}, {H{\'e}brard}, {Bonfils}, {Croll}, {Delfosse}, {Desort}, {Forveille},
  {Lagrange}, {Loeillet}, {Lovis}, {Matthews}, {Mayor}, {Pepe}, {Perrier},
  {Queloz}, {Rowe}, {Santos}, {S{\'e}gransan}, \& {Udry}}]{boisse09}
{Boisse}, I., {Moutou}, C., {Vidal-Madjar}, A., {et~al.} 2009, \aap, 495, 959

\bibitem[{{Bonfils} {et~al.}(2013){Bonfils}, {Delfosse}, {Udry}, {Forveille},
  {Mayor}, {Perrier}, {Bouchy}, {Gillon}, {Lovis}, {Pepe}, {Queloz}, {Santos},
  {S{\'e}gransan}, \& {Bertaux}}]{bonfils13}
{Bonfils}, X., {Delfosse}, X., {Udry}, S., {et~al.} 2013, \aap, 549, A109

\bibitem[{{Bus{\`a}} {et~al.}(2007){Bus{\`a}}, {Aznar Cuadrado}, {Terranegra},
  {Andretta}, \& {Gomez}}]{busa07}
{Bus{\`a}}, I., {Aznar Cuadrado}, R., {Terranegra}, L., {Andretta}, V., \&
  {Gomez}, M.~T. 2007, \aap, 466, 1089

\bibitem[{{Cenarro} {et~al.}(2001){Cenarro}, {Cardiel}, {Gorgas}, {Peletier},
  {Vazdekis}, \& {Prada}}]{cenarro01}
{Cenarro}, A.~J., {Cardiel}, N., {Gorgas}, J., {et~al.} 2001, \mnras, 326, 959

\bibitem[{{Chmielewski}(2000)}]{chmielewski00}
{Chmielewski}, Y. 2000, \aap, 353, 666

\bibitem[{{Cincunegui} {et~al.}(2007{\natexlab{a}}){Cincunegui}, {D{\'{\i}}az},
  \& {Mauas}}]{cincunegui07a}
{Cincunegui}, C., {D{\'{\i}}az}, R.~F., \& {Mauas}, P.~J.~D.
  2007{\natexlab{a}}, \aap, 461, 1107

\bibitem[{{Cincunegui} {et~al.}(2007{\natexlab{b}}){Cincunegui}, {D{\'{\i}}az},
  \& {Mauas}}]{cincunegui07b}
---. 2007{\natexlab{b}}, \aap, 469, 309

\bibitem[{{Clough} {et~al.}(2005){Clough}, {Shephard}, {Mlawer}, {Delamere},
  {Iacono}, {Cady-Pereira}, {Boukabara}, \& {Brown}}]{clough05}
{Clough}, S.~A., {Shephard}, M.~W., {Mlawer}, E.~J., {et~al.} 2005, \jqsrt, 91,
  233

\bibitem[{{Collins} {et~al.}(2016){Collins}, {Jones}, \& {Barnes}}]{collins16}
{Collins}, J.~M., {Jones}, H.~R.~A., \& {Barnes}, J.~R. 2016, ArXiv e-prints,
  arXiv:1608.07834

\bibitem[{{Davenport} {et~al.}(2016){Davenport}, {Kipping}, {Sasselov},
  {Matthews}, \& {Cameron}}]{davenport16}
{Davenport}, J.~R.~A., {Kipping}, D.~M., {Sasselov}, D., {Matthews}, J.~M., \&
  {Cameron}, C. 2016, ArXiv e-prints, arXiv:1608.06672

\bibitem[{{Dekker} {et~al.}(2000){Dekker}, {D'Odorico}, {Kaufer}, {Delabre}, \&
  {Kotzlowski}}]{dekker00}
{Dekker}, H., {D'Odorico}, S., {Kaufer}, A., {Delabre}, B., \& {Kotzlowski}, H.
  2000, in \procspie, Vol. 4008, Optical and IR Telescope Instrumentation and
  Detectors, ed. M.~{Iye} \& A.~F. {Moorwood}, 534--545

\bibitem[{{Deshpande} {et~al.}(2012){Deshpande}, {Mart{\'{\i}}n}, {Montgomery},
  {Zapatero Osorio}, {Rodler}, {del Burgo}, {Phan Bao}, {Lyubchik}, {Tata},
  {Bouy}, \& {Pavlenko}}]{deshpande12}
{Deshpande}, R., {Mart{\'{\i}}n}, E.~L., {Montgomery}, M.~M., {et~al.} 2012,
  \aj, 144, 99

\bibitem[{{D{\'{\i}}az} {et~al.}(2007){D{\'{\i}}az}, {Cincunegui}, \&
  {Mauas}}]{diaz07}
{D{\'{\i}}az}, R.~F., {Cincunegui}, C., \& {Mauas}, P.~J.~D. 2007, \mnras, 378,
  1007

\bibitem[{{Dumusque} {et~al.}(2015){Dumusque}, {Glenday}, {Phillips},
  {Buchschacher}, {Collier Cameron}, {Cecconi}, {Charbonneau}, {Cosentino},
  {Ghedina}, {Latham}, {Li}, {Lodi}, {Lovis}, {Molinari}, {Pepe}, {Udry},
  {Sasselov}, {Szentgyorgyi}, \& {Walsworth}}]{dumusque15}
{Dumusque}, X., {Glenday}, A., {Phillips}, D.~F., {et~al.} 2015, \apjl, 814,
  L21

\bibitem[{{Endl} {et~al.}(2006){Endl}, {Cochran}, {K{\"u}rster}, {Paulson},
  {Wittenmyer}, {MacQueen}, \& {Tull}}]{endl06}
{Endl}, M., {Cochran}, W.~D., {K{\"u}rster}, M., {et~al.} 2006, \apj, 649, 436

\bibitem[{{Endl} \& {K{\"u}rster}(2008)}]{endl08}
{Endl}, M., \& {K{\"u}rster}, M. 2008, \aap, 488, 1149

\bibitem[{{Endl} {et~al.}(2000){Endl}, {K{\"u}rster}, \& {Els}}]{endl00}
{Endl}, M., {K{\"u}rster}, M., \& {Els}, S. 2000, \aap, 362, 585

\bibitem[{{Fischer} {et~al.}(2016){Fischer}, {Anglada-Escude}, {Arriagada},
  {Baluev}, {Bean}, {Bouchy}, {Buchhave}, {Carroll}, {Chakraborty}, {Crepp},
  {Dawson}, {Diddams}, {Dumusque}, {Eastman}, {Endl}, {Figueira}, {Ford},
  {Foreman-Mackey}, {Fournier}, {F{\H u}r{\'e}sz}, {Gaudi}, {Gregory},
  {Grundahl}, {Hatzes}, {H{\'e}brard}, {Herrero}, {Hogg}, {Howard}, {Johnson},
  {Jorden}, {Jurgenson}, {Latham}, {Laughlin}, {Loredo}, {Lovis}, {Mahadevan},
  {McCracken}, {Pepe}, {Perez}, {Phillips}, {Plavchan}, {Prato}, {Quirrenbach},
  {Reiners}, {Robertson}, {Santos}, {Sawyer}, {Segransan}, {Sozzetti},
  {Steinmetz}, {Szentgyorgyi}, {Udry}, {Valenti}, {Wang}, {Wittenmyer}, \&
  {Wright}}]{fischer16}
{Fischer}, D.~A., {Anglada-Escude}, G., {Arriagada}, P., {et~al.} 2016, \pasp,
  128, 066001

\bibitem[{{Fuhrmeister} {et~al.}(2011){Fuhrmeister}, {Lalitha}, {Poppenhaeger},
  {Rudolf}, {Liefke}, {Reiners}, {Schmitt}, \& {Ness}}]{fuhrmeister11}
{Fuhrmeister}, B., {Lalitha}, S., {Poppenhaeger}, K., {et~al.} 2011, \aap, 534,
  A133

\bibitem[{{Gillon} {et~al.}(2016){Gillon}, {Jehin}, {Lederer}, {Delrez}, {de
  Wit}, {Burdanov}, {Van Grootel}, {Burgasser}, {Triaud}, {Opitom}, {Demory},
  {Sahu}, {Bardalez Gagliuffi}, {Magain}, \& {Queloz}}]{gillon16}
{Gillon}, M., {Jehin}, E., {Lederer}, S.~M., {et~al.} 2016, \nat, 533, 221

\bibitem[{{Gomes da Silva} {et~al.}(2011){Gomes da Silva}, {Santos}, {Bonfils},
  {Delfosse}, {Forveille}, \& {Udry}}]{gds11}
{Gomes da Silva}, J., {Santos}, N.~C., {Bonfils}, X., {et~al.} 2011, \aap, 534,
  A30

\bibitem[{{Hanuschik}(2003)}]{hanuschik03}
{Hanuschik}, R.~W. 2003, \aap, 407, 1157

\bibitem[{{Hatzes} {et~al.}(2015){Hatzes}, {Cochran}, {Endl}, {Guenther},
  {MacQueen}, {Hartmann}, {Zechmeister}, {Han}, {Lee}, {Walker}, {Yang},
  {Larson}, {Kim}, {Mkrtichian}, {D{\"o}llinger}, {Simon}, \&
  {Girardi}}]{hatzes15}
{Hatzes}, A.~P., {Cochran}, W.~D., {Endl}, M., {et~al.} 2015, \aap, 580, A31

\bibitem[{{Hawley} \& {Pettersen}(1991)}]{hawley91}
{Hawley}, S.~L., \& {Pettersen}, B.~R. 1991, \apj, 378, 725

\bibitem[{{Haywood} {et~al.}(2014){Haywood}, {Collier Cameron}, {Queloz},
  {Barros}, {Deleuil}, {Fares}, {Gillon}, {Lanza}, {Lovis}, {Moutou}, {Pepe},
  {Pollacco}, {Santerne}, {S{\'e}gransan}, \& {Unruh}}]{haywood14}
{Haywood}, R.~D., {Collier Cameron}, A., {Queloz}, D., {et~al.} 2014, \mnras,
  443, 2517

\bibitem[{{Haywood} {et~al.}(2016){Haywood}, {Collier Cameron}, {Unruh},
  {Lovis}, {Lanza}, {Llama}, {Deleuil}, {Fares}, {Gillon}, {Moutou}, {Pepe},
  {Pollacco}, {Queloz}, \& {S{\'e}gransan}}]{haywood16}
{Haywood}, R.~D., {Collier Cameron}, A., {Unruh}, Y.~C., {et~al.} 2016, \mnras,
  457, 3637

\bibitem[{{Howard} {et~al.}(2013){Howard}, {Sanchis-Ojeda}, {Marcy}, {Johnson},
  {Winn}, {Isaacson}, {Fischer}, {Fulton}, {Sinukoff}, \& {Fortney}}]{howard13}
{Howard}, A.~W., {Sanchis-Ojeda}, R., {Marcy}, G.~W., {et~al.} 2013, \nat, 503,
  381

\bibitem[{{Isaacson} \& {Fischer}(2010)}]{if10}
{Isaacson}, H., \& {Fischer}, D. 2010, \apj, 725, 875

\bibitem[{{Johnson} {et~al.}(2016){Johnson}, {Endl}, {Cochran}, {Meschiari},
  {Robertson}, {MacQueen}, {Brugamyer}, {Caldwell}, {Hatzes}, {Ram{\'{\i}}rez},
  \& {Wittenmyer}}]{johnson16}
{Johnson}, M.~C., {Endl}, M., {Cochran}, W.~D., {et~al.} 2016, ArXiv e-prints,
  arXiv:1602.05200

\bibitem[{{Kafka} \& {Honeycutt}(2006)}]{kafka06}
{Kafka}, S., \& {Honeycutt}, R.~K. 2006, \aj, 132, 1517

\bibitem[{{Kiraga} \& {Stepien}(2007)}]{kiraga07}
{Kiraga}, M., \& {Stepien}, K. 2007, \actaa, 57, 149

\bibitem[{{Kowalski} {et~al.}(2013){Kowalski}, {Hawley}, {Wisniewski}, {Osten},
  {Hilton}, {Holtzman}, {Schmidt}, \& {Davenport}}]{kowalski13}
{Kowalski}, A.~F., {Hawley}, S.~L., {Wisniewski}, J.~P., {et~al.} 2013, \apjs,
  207, 15

\bibitem[{{K{\"u}rster} {et~al.}(2003){K{\"u}rster}, {Endl}, {Rouesnel}, {Els},
  {Kaufer}, {Brillant}, {Hatzes}, {Saar}, \& {Cochran}}]{kurster03}
{K{\"u}rster}, M., {Endl}, M., {Rouesnel}, F., {et~al.} 2003, \aap, 403, 1077

\bibitem[{{Linsky} {et~al.}(1970){Linsky}, {Teske}, \& {Wilkinson}}]{linsky70}
{Linsky}, J.~L., {Teske}, R.~G., \& {Wilkinson}, C.~W. 1970, \solphys, 11, 374

\bibitem[{{Lockwood} {et~al.}(2014){Lockwood}, {Johnson}, {Bender}, {Carr},
  {Barman}, {Richert}, \& {Blake}}]{lockwood14}
{Lockwood}, A.~C., {Johnson}, J.~A., {Bender}, C.~F., {et~al.} 2014, \apjl,
  783, L29

\bibitem[{{Lovis} {et~al.}(2011){Lovis}, {Dumusque}, {Santos}, {Bouchy},
  {Mayor}, {Pepe}, {Queloz}, {S{\'e}gransan}, \& {Udry}}]{lovis11}
{Lovis}, C., {Dumusque}, X., {Santos}, N.~C., {et~al.} 2011, ArXiv e-prints,
  arXiv:1107.5325

\bibitem[{{Mahadevan} {et~al.}(2014){Mahadevan}, {Ramsey}, {Terrien},
  {Halverson}, {Roy}, {Hearty}, {Levi}, {Stefansson}, {Robertson}, {Bender},
  {Schwab}, \& {Nelson}}]{mahadevan14}
{Mahadevan}, S., {Ramsey}, L.~W., {Terrien}, R., {et~al.} 2014, in \procspie,
  Vol. 9147, Ground-based and Airborne Instrumentation for Astronomy V, 91471G

\bibitem[{{Mallik}(1997)}]{mallik97}
{Mallik}, S.~V. 1997, \aaps, 124, doi:10.1051/aas:1997199

\bibitem[{{Marchwinski} {et~al.}(2015){Marchwinski}, {Mahadevan}, {Robertson},
  {Ramsey}, \& {Harder}}]{marchwinski15}
{Marchwinski}, R.~C., {Mahadevan}, S., {Robertson}, P., {Ramsey}, L., \&
  {Harder}, J. 2015, \apj, 798, 63

\bibitem[{{Mortier} {et~al.}(2016){Mortier}, {Faria}, {Santos}, {Rajpaul},
  {Figueira}, {Boisse}, {Collier Cameron}, {Dumusque}, {Lo Curto}, {Lovis},
  {Mayor}, {Melo}, {Pepe}, {Queloz}, {Santerne}, {S{\'e}gransan}, {Sousa},
  {Sozzetti}, \& {Udry}}]{mortier16}
{Mortier}, A., {Faria}, J.~P., {Santos}, N.~C., {et~al.} 2016, \aap, 585, A135

\bibitem[{{Pepe} {et~al.}(2000){Pepe}, {Mayor}, {Delabre}, {Kohler}, {Lacroix},
  {Queloz}, {Udry}, {Benz}, {Bertaux}, \& {Sivan}}]{pepe00}
{Pepe}, F., {Mayor}, M., {Delabre}, B., {et~al.} 2000, in \procspie, Vol. 4008,
  Optical and IR Telescope Instrumentation and Detectors, ed. M.~{Iye} \& A.~F.
  {Moorwood}, 582--592

\bibitem[{{Pepe} {et~al.}(2004){Pepe}, {Mayor}, {Queloz}, {Benz}, {Bonfils},
  {Bouchy}, {Lo Curto}, {Lovis}, {M{\'e}gevand}, {Moutou}, {Naef}, {Rupprecht},
  {Santos}, {Sivan}, {Sosnowska}, \& {Udry}}]{pepe04}
{Pepe}, F., {Mayor}, M., {Queloz}, D., {et~al.} 2004, \aap, 423, 385

\bibitem[{{Pepe} {et~al.}(2013){Pepe}, {Cameron}, {Latham}, {Molinari}, {Udry},
  {Bonomo}, {Buchhave}, {Charbonneau}, {Cosentino}, {Dressing}, {Dumusque},
  {Figueira}, {Fiorenzano}, {Gettel}, {Harutyunyan}, {Haywood}, {Horne},
  {Lopez-Morales}, {Lovis}, {Malavolta}, {Mayor}, {Micela}, {Motalebi},
  {Nascimbeni}, {Phillips}, {Piotto}, {Pollacco}, {Queloz}, {Rice}, {Sasselov},
  {S{\'e}gransan}, {Sozzetti}, {Szentgyorgyi}, \& {Watson}}]{pepe13}
{Pepe}, F., {Cameron}, A.~C., {Latham}, D.~W., {et~al.} 2013, \nat, 503, 377

\bibitem[{{Pojmanski}(1997)}]{pojmanski97}
{Pojmanski}, G. 1997, \actaa, 47, 467

\bibitem[{{Queloz} {et~al.}(2001){Queloz}, {Henry}, {Sivan}, {Baliunas},
  {Beuzit}, {Donahue}, {Mayor}, {Naef}, {Perrier}, \& {Udry}}]{queloz01}
{Queloz}, D., {Henry}, G.~W., {Sivan}, J.~P., {et~al.} 2001, \aap, 379, 279

\bibitem[{{Quirrenbach} {et~al.}(2014){Quirrenbach}, {Amado}, {Caballero},
  {Mundt}, {Reiners}, {Ribas}, {Seifert}, {Abril}, {Aceituno},
  {Alonso-Floriano}, {Ammler-von Eiff}, {Antona Jim{\'e}nez},
  {Anwand-Heerwart}, {Azzaro}, {Bauer}, {Barrado}, {Becerril}, {B{\'e}jar},
  {Ben{\'{\i}}tez}, {Berdi{\~n}as}, {C{\'a}rdenas}, {Casal}, {Claret},
  {Colom{\'e}}, {Cort{\'e}s-Contreras}, {Czesla}, {Doellinger}, {Dreizler},
  {Feiz}, {Fern{\'a}ndez}, {Galad{\'{\i}}}, {G{\'a}lvez-Ortiz},
  {Garc{\'{\i}}a-Piquer}, {Garc{\'{\i}}a-Vargas}, {Garrido}, {Gesa}, {G{\'o}mez
  Galera}, {Gonz{\'a}lez {\'A}lvarez}, {Gonz{\'a}lez Hern{\'a}ndez},
  {Gr{\"o}zinger}, {Gu{\`a}rdia}, {Guenther}, {de Guindos},
  {Guti{\'e}rrez-Soto}, {Hagen}, {Hatzes}, {Hauschildt}, {Helmling}, {Henning},
  {Hermann}, {Hern{\'a}ndez Casta{\~n}o}, {Herrero}, {Hidalgo}, {Holgado},
  {Huber}, {Huber}, {Jeffers}, {Joergens}, {de Juan}, {Kehr}, {Klein},
  {K{\"u}rster}, {Lamert}, {Lalitha}, {Laun}, {Lemke}, {Lenzen}, {L{\'o}pez del
  Fresno}, {L{\'o}pez Mart{\'{\i}}}, {L{\'o}pez-Santiago}, {Mall}, {Mandel},
  {Mart{\'{\i}}n}, {Mart{\'{\i}}n-Ruiz}, {Mart{\'{\i}}nez-Rodr{\'{\i}}guez},
  {Marvin}, {Mathar}, {Mirabet}, {Montes}, {Morales Mu{\~n}oz}, {Moya},
  {Naranjo}, {Ofir}, {Oreiro}, {Pall{\'e}}, {Panduro}, {Passegger},
  {P{\'e}rez-Calpena}, {P{\'e}rez Medialdea}, {Perger}, {Pluto}, {Ram{\'o}n},
  {Rebolo}, {Redondo}, {Reffert}, {Reinhardt}, {Rhode}, {Rix}, {Rodler},
  {Rodr{\'{\i}}guez}, {Rodr{\'{\i}}guez-L{\'o}pez},
  {Rodr{\'{\i}}guez-P{\'e}rez}, {Rohloff}, {Rosich}, {S{\'a}nchez-Blanco},
  {S{\'a}nchez Carrasco}, {Sanz-Forcada}, {Sarmiento}, {Sch{\"a}fer},
  {Schiller}, {Schmidt}, {Schmitt}, {Solano}, {Stahl}, {Storz}, {St{\"u}rmer},
  {Su{\'a}rez}, {Ulbrich}, {Veredas}, {Wagner}, {Winkler}, {Zapatero Osorio},
  {Zechmeister}, {Abell{\'a}n de Paco}, {Anglada-Escud{\'e}}, {del Burgo},
  {Klutsch}, {Lizon}, {L{\'o}pez-Morales}, {Morales}, {Perryman}, {Tulloch}, \&
  {Xu}}]{quirrenbach14}
{Quirrenbach}, A., {Amado}, P.~J., {Caballero}, J.~A., {et~al.} 2014, in
  \procspie, Vol. 9147, Ground-based and Airborne Instrumentation for Astronomy
  V, 91471F

\bibitem[{{Reiners}(2009)}]{reiners09}
{Reiners}, A. 2009, \aap, 498, 853

\bibitem[{{Robertson} {et~al.}(2013){Robertson}, {Endl}, {Cochran}, \&
  {Dodson-Robinson}}]{robertson13}
{Robertson}, P., {Endl}, M., {Cochran}, W.~D., \& {Dodson-Robinson}, S.~E.
  2013, \apj, 764, 3

\bibitem[{{Robertson} {et~al.}(2014){Robertson}, {Mahadevan}, {Endl}, \&
  {Roy}}]{robertson14}
{Robertson}, P., {Mahadevan}, S., {Endl}, M., \& {Roy}, A. 2014, Science, 345,
  440

\bibitem[{{Robertson} {et~al.}(2015){Robertson}, {Roy}, \&
  {Mahadevan}}]{robertson15}
{Robertson}, P., {Roy}, A., \& {Mahadevan}, S. 2015, \apjl, 805, L22

\bibitem[{{Rothman} {et~al.}(2009){Rothman}, {Gordon}, {Barbe}, {Benner},
  {Bernath}, {Birk}, {Boudon}, {Brown}, {Campargue}, {Champion}, {Chance},
  {Coudert}, {Dana}, {Devi}, {Fally}, {Flaud}, {Gamache}, {Goldman},
  {Jacquemart}, {Kleiner}, {Lacome}, {Lafferty}, {Mandin}, {Massie},
  {Mikhailenko}, {Miller}, {Moazzen-Ahmadi}, {Naumenko}, {Nikitin}, {Orphal},
  {Perevalov}, {Perrin}, {Predoi-Cross}, {Rinsland}, {Rotger}, {{\v S}ime{\v
  c}kov{\'a}}, {Smith}, {Sung}, {Tashkun}, {Tennyson}, {Toth}, {Vandaele}, \&
  {Vander Auwera}}]{rothman09}
{Rothman}, L.~S., {Gordon}, I.~E., {Barbe}, A., {et~al.} 2009, \jqsrt, 110, 533

\bibitem[{{Santos} {et~al.}(2014){Santos}, {Mortier}, {Faria}, {Dumusque},
  {Adibekyan}, {Delgado-Mena}, {Figueira}, {Benamati}, {Boisse}, {Cunha},
  {Gomes da Silva}, {Lo Curto}, {Lovis}, {Martins}, {Mayor}, {Melo}, {Oshagh},
  {Pepe}, {Queloz}, {Santerne}, {S{\'e}gransan}, {Sozzetti}, {Sousa}, \&
  {Udry}}]{santos14}
{Santos}, N.~C., {Mortier}, A., {Faria}, J.~P., {et~al.} 2014, \aap, 566, A35

\bibitem[{{Schlieder} {et~al.}(2012){Schlieder}, {L{\'e}pine}, {Rice}, {Simon},
  {Fielding}, \& {Tomasino}}]{schlieder12}
{Schlieder}, J.~E., {L{\'e}pine}, S., {Rice}, E., {et~al.} 2012, \aj, 143, 114

\bibitem[{{Schmidt} {et~al.}(2012){Schmidt}, {Kowalski}, {Hawley}, {Hilton},
  {Wisniewski}, \& {Tofflemire}}]{schmidt12}
{Schmidt}, S.~J., {Kowalski}, A.~F., {Hawley}, S.~L., {et~al.} 2012, \apj, 745,
  14

\bibitem[{{Seifahrt} {et~al.}(2016){Seifahrt}, {Bean}, {St{\"u}rmer}, {Gers},
  {Grobler}, {Reed}, \& {Jones}}]{seifahrt16}
{Seifahrt}, A., {Bean}, J.~L., {St{\"u}rmer}, J., {et~al.} 2016, ArXiv
  e-prints, arXiv:1606.07140

\bibitem[{{Su{\'a}rez Mascare{\~n}o} {et~al.}(2016){Su{\'a}rez Mascare{\~n}o},
  {Rebolo}, \& {Gonz{\'a}lez Hern{\'a}ndez}}]{suarez16}
{Su{\'a}rez Mascare{\~n}o}, A., {Rebolo}, R., \& {Gonz{\'a}lez Hern{\'a}ndez},
  J.~I. 2016, ArXiv e-prints, arXiv:1607.03049

\bibitem[{{Su{\'a}rez Mascare{\~n}o} {et~al.}(2015){Su{\'a}rez Mascare{\~n}o},
  {Rebolo}, {Gonz{\'a}lez Hern{\'a}ndez}, \& {Esposito}}]{suarez15}
{Su{\'a}rez Mascare{\~n}o}, A., {Rebolo}, R., {Gonz{\'a}lez Hern{\'a}ndez},
  J.~I., \& {Esposito}, M. 2015, \mnras, 452, 2745

\bibitem[{{Takeda} {et~al.}(2002){Takeda}, {Zhao}, {Chen}, {Qiu}, \&
  {Takada-Hidai}}]{takeda02}
{Takeda}, Y., {Zhao}, G., {Chen}, Y.-Q., {Qiu}, H.-M., \& {Takada-Hidai}, M.
  2002, \pasj, 54, 275

\bibitem[{{Terrien} {et~al.}(2015){Terrien}, {Mahadevan}, {Bender},
  {Deshpande}, \& {Robertson}}]{terrien15}
{Terrien}, R.~C., {Mahadevan}, S., {Bender}, C.~F., {Deshpande}, R., \&
  {Robertson}, P. 2015, \apjl, 802, L10

\bibitem[{{Torres} {et~al.}(2006){Torres}, {Quast}, {da Silva}, {de La Reza},
  {Melo}, \& {Sterzik}}]{torres06}
{Torres}, C.~A.~O., {Quast}, G.~R., {da Silva}, L., {et~al.} 2006, \aap, 460,
  695

\bibitem[{{Valenti} {et~al.}(1995){Valenti}, {Butler}, \& {Marcy}}]{valenti95}
{Valenti}, J.~A., {Butler}, R.~P., \& {Marcy}, G.~W. 1995, \pasp, 107, 966

\bibitem[{{Vaughan} {et~al.}(1978){Vaughan}, {Preston}, \&
  {Wilson}}]{vaughan78}
{Vaughan}, A.~H., {Preston}, G.~W., \& {Wilson}, O.~C. 1978, \pasp, 90, 267

\bibitem[{{Wade} \& {Horne}(1988)}]{wade88}
{Wade}, R.~A., \& {Horne}, K. 1988, \apj, 324, 411

\bibitem[{{West} {et~al.}(2015){West}, {Weisenburger}, {Irwin},
  {Berta-Thompson}, {Charbonneau}, {Dittmann}, \& {Pineda}}]{west15}
{West}, A.~A., {Weisenburger}, K.~L., {Irwin}, J., {et~al.} 2015, \apj, 812, 3

\bibitem[{{Wilson}(1968)}]{wilson68}
{Wilson}, O.~C. 1968, \apj, 153, 221

\bibitem[{{Wright}(2005)}]{wright05}
{Wright}, J.~T. 2005, \pasp, 117, 657

\bibitem[{{Zechmeister} \& {K{\"u}rster}(2009)}]{zk09}
{Zechmeister}, M., \& {K{\"u}rster}, M. 2009, \aap, 496, 577

\bibitem[{{Zechmeister} {et~al.}(2009){Zechmeister}, {K{\"u}rster}, \&
  {Endl}}]{zechmeister09}
{Zechmeister}, M., {K{\"u}rster}, M., \& {Endl}, M. 2009, \aap, 505, 859

\end{thebibliography}

\clearpage

\end{document}